\documentclass[final,12pt]{elsarticle}

\usepackage{hyperref}

\usepackage{color}
\definecolor{darkblue}{rgb}{0.,0.,0.4}
\definecolor{darkred}{rgb}{0.5,0.,0.}

\usepackage{graphicx} 
\usepackage{revsymb}
\usepackage{bm}
\usepackage{textcomp}
\usepackage{revsymb}
\usepackage{amssymb}
\usepackage{ifsym}
\usepackage{wasysym}
\usepackage{txfonts}
\usepackage{tipa}
\usepackage{cleveref}[2012/02/15]   
\crefformat{footnote}{#2\footnotemark[#1]#3}
\usepackage{multirow}

\begin{document}
	
	\begin{frontmatter}

	\title{Identification of CaCuSi$_4$O$_{10}$ (Egyptian blue) in the ``Birch. Spring'' painting by Robert Falk (1907) using photoluminescence%
}

	

		
		
		\author[GOSNIIR]{Svetlana~A.~Pisareva}
		\ead{pisarevasa@gosniir.ru}
		
		\author[Tretyakov_Expertise]{Irina~N.~Shibanova}
		\ead{shibanovacv@gmail.com}
		

		\author[GOSNIIR]{Irina~F.~Kadikova\corref{correspondingauthor}} 
		\ead{kadikovaif@gosniir.ru}

		\author[GOSNIIR]{Ekaterina~A.~Morozova}
		\ead{morozovaea@gosniir.ru}
		
		\author[GOSNIIR]{Tatyana~V.~Yuryeva} 
		\ead{yuryevatv@gosniir.ru}
		
		\author[RF_CFS]{Ilya~B.~Afanasyev}
		\ead{il.afanasyev@sudexpert.ru}
		
		
		\author[GPI]{Vladimir~A.~Yuryev\corref{correspondingauthor}}
		\ead{vyuryev@kapella.gpi.ru}

		\address[GOSNIIR]{The State Research Institute for Restoration, Building~1, 44~Gastello Street, Moscow 107114, Russia}
		
		\address[Tretyakov_Expertise]{P.\,M.\,Tretyakov Scientific-Research Independent Expert Facility, Ltd., Building~1, 44~Gastello Street, Moscow 107114, Russia}

		
		
		
		
		\address[RF_CFS]{The Russian Federal Center of Forensic Science of the Ministry of Justice, Building~2, 13~Khokhlovskiy Sidestreet, Moscow 109028, Russia}
		
		
		\address[GPI]{A.\,M.\,Prokhorov General Physics Institute of the Russian Academy of Sciences, 38~Vavilov Street, Moscow 119991, Russia}
		

		\date{ }

\cortext[correspondingauthor]{Corresponding author}


\begin{abstract}
\footnotesize 
We have detected Egyptian blue pigment in the paint layer of the ``Birch. Spring'' painting by~Robert~Falk (1907); 
we have also found this pigment in the paints of the sketch drawn on the canvas back side. 
This is probably the first discovery of Egyptian blue in a 20th century work of art.	
We have analyzed a modern commercial Egyptian blue pigment (Kremer) and found it to be suitable as a standard for photoluminescence spectral analysis.
The characteristic photoluminescence band of CaCuSi$_4$O$_{10}$ reaches maximum at the wavelength of about 910~nm.
The luminescence is efficiently excited by incoherent green or blue light.
The study demonstrates that the photoluminescence spectral micro analysis using excitation by incoherent light can be efficiently used for the identification of luminescent pigments in paint layers of artworks.
\end{abstract}




\begin{keyword}\footnotesize
	\texttt{Egyptian blue \sep Photoluminescence \sep Polarizing microscopy \sep Elemental mapping \sep Paints analysis \sep  Russian avant-garde painting}
\end{keyword}

\end{frontmatter}


\newpage

\section*{Graphical abstract}
\begin{figure}[h]
	\begin{minipage}[l]{1\textwidth}
		\includegraphics[width=0.95\textwidth]{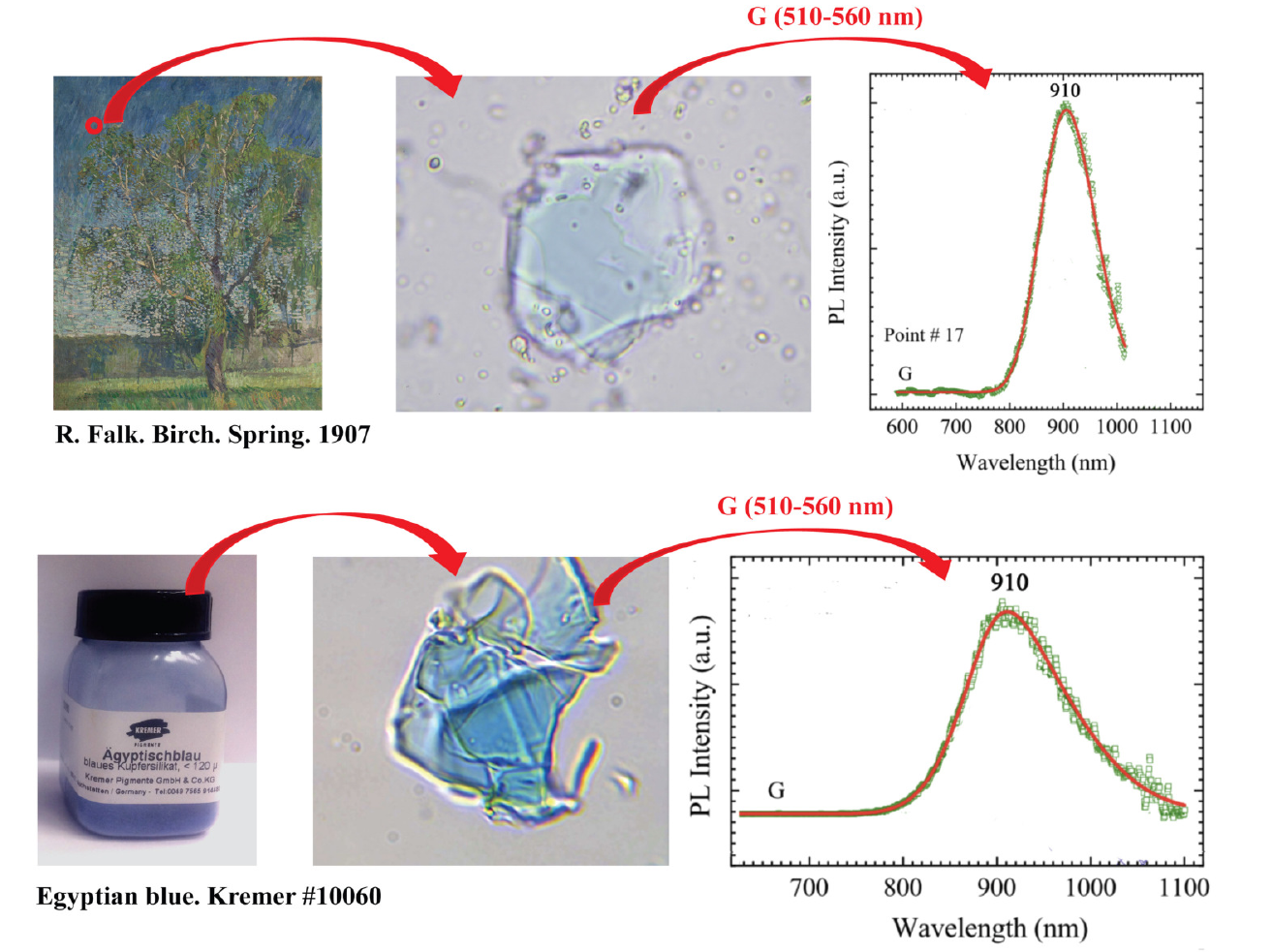}
	\end{minipage}
	\label{fig:grabs}
\end{figure}

\newpage

\section*{Research highlights}
\begin{itemize}
\item Egyptian blue is identified in ``Birch. Spring'' painting by R.~Falk (1907)
\item CaCuSi$_4$O$_{10}$ emission band is centered at the wavelength of 910~nm
\item The luminescence is efficiently excited by incoherent green or blue light
\item Luminescence enables detecting some pigments in paint layers of artworks
\item Commercial Egyptian blue (Kremer) fits as a standard for spectral analysis
\end{itemize}

\newpage
\section{Introduction}
\label{intr}

\subsection{Egyptian blue from the antiquity to the present: a brief historical reference}
\label{history}

Egyptian blue, also known as calcium copper tetrasilicate or calcium copper phyllo-tetrasilicate (CaCuSi$_4$O$_{10}$), {Alexandria frit} or {Pompeian blue}, is one of the earliest synthetic pigments first obtained in Egypt no later than the 4th dynasty (ca.~2613--2494 BC) \cite{Pigment_Compendium,Egyptian_Blue_book,Ancient_Blue_Purple_Pigments} and widely used in the antiquity \cite{Pigment_Compendium,Micro-EB_Arch_England,EB_Europe_Egypt-1910} by different names \cite{Egyptian_Blue_book}.%
\footnote{%
	Remind that, in antiquity, owing to the imperfectness of the production processes, Egyptian blue pigment was frit that, apart from calcium copper tetrasilicate, comprised appreciable quantities of numerous impurities such as, e.g., 
	wollastonite (CaSiO$_3$), Cu-rich glass, cuprite (Cu$_2$O), tenorite (CuO) \cite{Egyptian_Blue_Fayum_Portraits-2015}, cassiterite (SnO$_2$), malayaite (CaSnSiO$_5$), quartz and free lime  \cite{EB_production_technology_Mesopotamia}.
	At present, the term of Egyptian blue usually relates to CaCuSi$_4$O$_{10}$ itself, especially if the art pigment is concerned.
	}
	Ancient Romans, e.g., knew this substance by the name \textit{Caeruleum aegyptium} \cite{Egyptian_Blue_book,Colour_Atlas,arrigoni2015colour,diodato2017encaustic}.
	Also, it was extensively produced throughout ancient Western Asia and the Mediterranean Region (Mesopotamia, Persia, Assyria, Urartu, Parthia and Greece) \cite{Egyptian_Blue_book,Ancient_Blue_Purple_Pigments,EB_production_technology_Mesopotamia,Egyptian_Blue_Persia,EB_Turkey_Ayanis,EB_Turkey_Lake_Van,EB_Egypt_Aegean+Near_East,EB@Kos_article,Grenberg_&_Pisareva-Erebuni_1982,Pisareva-Erebuni_1987,Egyptian_Blue_Visible-Induced,Egyptian_Blue_Fayum_Portraits-2015,Egyptian_Blue_Fayum_Portraits-2018,Old_Nisa_Veresotskaya}.
	It was found out as one of the main blue pigment in the Fayum mummy portraits (1st to 3d centuries AD) \cite{Egyptian_Blue_Fayum_Portraits-2015,Egyptian_Blue_Fayum_Portraits-2018} 
	and in the paint layer of the early Christian encaustic icons ``Male and Female Martyrs'' and ``Sergius and Bacchus'' 
	\cite{EB_icons_Naumova-1983}.
	By the 3rd century, Egyptian blue spread in the Roman Europe (see Ref.~\cite{Micro-EB_Arch_England} and references therein) and even in Norway \cite{EB_Norway}. %

Egyptian blue related to the early medieval period was reported to be detected in the mural painting of a church of the 9th or 10th century AD (Lombardy, Italy) \cite{EB_Zn-rich_Nicola2019,EB_Zn-rich_Castelseprio_NICOLA2018465}
and 
in an 11th century mural altarpiece of the church of Sant Pere (Terrassa, Spain)
\cite{EB_Spain}.
Later, in the Middle Ages, Egyptian blue, as considered, was lost \cite{Egyptian_Blue_book}.
However, CaCuSi$_4$O$_{10}$ was unexpectedly identified using optical and scanning electron microscopy, photoluminescence, and Raman microspectroscopy in the  painting of St.~Margaret by Ortolano Ferrarese (Giovanni Battista Benvenuto) made in Italy in 1524~\cite{Egyptian_Blue_Benvenuto-1524}.

The natural form of calcium copper tetrasilicate, cuprorivaite, was described as a new mineral from Vesuvius in 1938 \cite{Cuprorivaite_discovery}.
Its structure is presented, e.g., in Refs.~\cite{Pabst:cuprorivaite,Reexamination_cuprorivaite,Cuprorivaite_Structure,Bensch:cuprorivaite,Cuprorivaite,Cuprorivaite_mindata}.
However, it is a rarely met mineral usually intimately mixed with quartz \cite{Cuprorivaite_mindata}; it is mainly distributed in Italy, Germany, USA and South Africa \cite{Cuprorivaite,Cuprorivaite_mindata}.

Nowadays, calcium copper tetrasilicate, due to its unique luminescence characteristics  \cite{Egyptian_Blue_PL,PL_some_Blue_natural_pigments,EB-PL_Application_Triolo2019,MCuSi4O10_IR_phosphors,EB_NIR_emission_upconversion,PL_EB_HB_HP}, has achieved increasing importance in science and technology \cite{Egyptian_Blue_New_Life,Warner_EB_Synthesis-2011,Egyptian_Blue_Nano,EB_hydrothermal,EB_Exfoliation,EB_ML_structure&properties,EB_fingerprint_powder}; it may find numerous applications in various areas of photonics soon.

\begin{figure}[t]
	\includegraphics[scale=0.4]{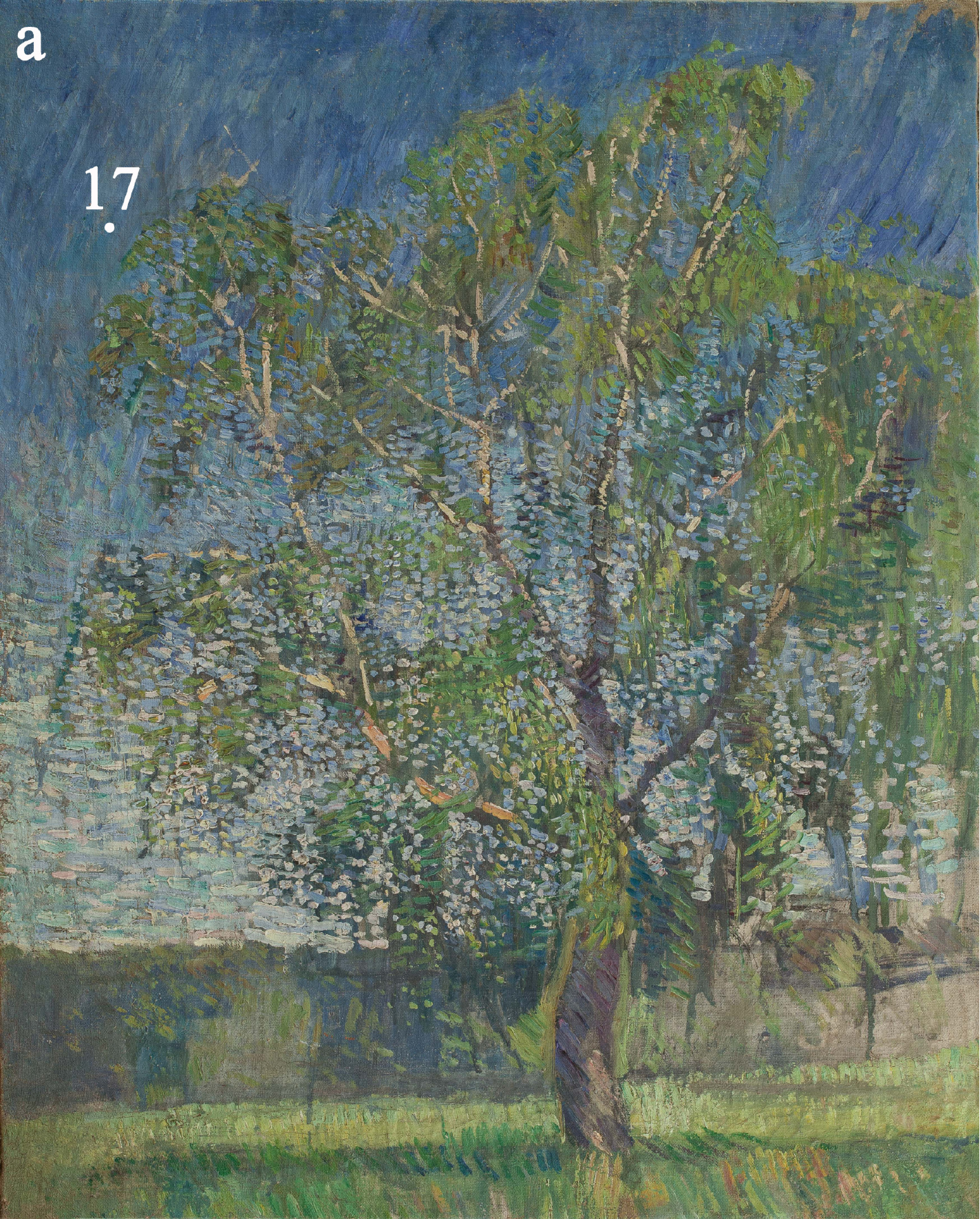}
	\includegraphics[scale=0.4]{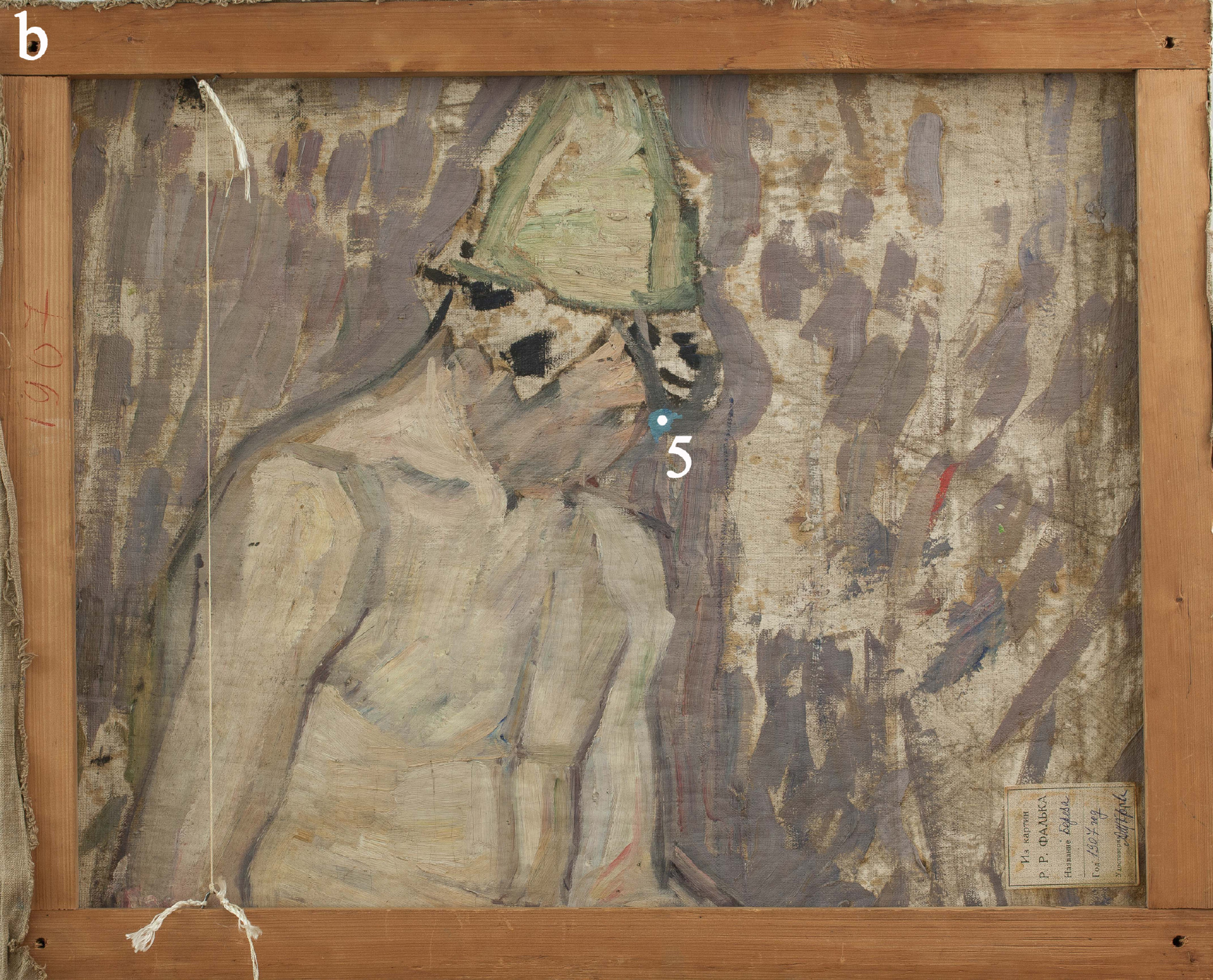}%
	\caption{%
		Robert Falk. ``Birch. Spring.'' 1907. Oil on canvas. $79\times63$ cm. Private collection.
		Photographs of the front (a) and rear (b) sides; the numbered filled circles indicate the points, at which the studied samples were taken (\#\,5 and \#\,17).
		\label{fig:Falk_painting}
	}
\end{figure}

\subsection{On the ``Birch. Spring'' painting by~Robert~Falk}
\label{Falk_Intr}

Surprisingly, calcium copper tetrasilicate has recently been detected using energy dispersive X-ray spectroscopy and polarizing microscopy in the 20th century painting ``Birch. Spring'' (oil on canvas, private collection, see Fig.~\ref{fig:Falk_painting}) \cite{R.Falk_Pisareva}.

The author of the picture, Robert Falk (1886--1956) is one of the bright representatives of Russian avant-garde, a leading member of ``The Jack of Diamonds'' creative association (1910--1916) \cite{Falk_Besancon,Falk_Sarabjanow,Falk_Stschekin-Krotowa,My_Falk_Stschekin-Krotowa}. 
However, the painting dates back to the early period of Robert Falk's work, the time of his formation as an artist, and was painted in his second year at the Moscow School of Painting, Sculpture and Architecture in 1907 \cite{Falk_catalog}. 

A two-sided image on paintings is a characteristic feature of the artist's work at that time. 
Repeatedly, in need of a clean canvas, the artist turned over his work (and cut some into several parts) and wrote down the clean back of the canvas. 
This is exactly what he did with the ``Birch. Spring'' painting, when depicting a sketch of a human figure on the back of the canvas (Fig.~\ref{fig:Falk_painting}).
Interestingly, that Egyptian blue was identified in the blue paint layers of both images, which indirectly supports the assumption about close time of creation of both the spring landscape and the figured sketch on the back side of the canvas. 



This article presents a detailed analysis of two paint samples (Fig.~\ref{fig:Falk_painting}), obtained from this painting, performed using a number of complementary analytical techniques, which has completely verified the previously reported observation of CaCuSi$_4$O$_{10}$ in blue paints of this picture \cite{R.Falk_Pisareva}.

\section{Experimental details}
\label{exp}

\subsection{Methods and equipment}
\label{methods}


We have applied the multi-analytical approach to both the characterization of the commercial Egyptian blue pigment (Kremer), used as a standard in this work, and the studies of the paint layer composition of  the ``Birch. Spring'' painting by~Robert~Falk presented in this article. 
The following techniques and equipment were used in the research.

Mira 3~XMU (Tescan Orsay Holding) scanning electron microscope (SEM) was employed for sample imaging and elemental analysis.
Pairs of SEM images were recorded simultaneously using secondary (SE) and backscattered (BSE) electrons.%
\footnote{%
	As a rule, we record pairs of SEM SE and BSE images at each point since SEM running in the former mode it renders mainly the spatial relief of the studied surface, while operated in the latter mode it demonstrates mainly the substance density (i.e. its elemental or phase composition), whereas the information on both relief and composition of a sample is usually required for the correct interpretation of data obtained using SEM.
}
Elemental micro\-analysis and mapping was made using
EDS X-MAX 50 (Oxford Instruments Nanoanalysis) energy dispersive X-ray (EDX) spectrometer.

Binder was analyzed using Tensor 27 Fourier-transform infrared spectrometer (Bruker) equipped with MIRacle attenuated total reflection (ATR) accessory (PIKE Technologies) with a ZnSe ATR crystal.

X-ray phase analysis was carried out by means of the Debye-Scherrer powder diffraction \cite{X-ray_Diffraction_Book} using D8 Advance diffractometer (Bruker) at non-monochro\-matic Cu\,K$_{\alpha}$ band (Cu\,K$_{\alpha_{1,2}}$, $\lambdaup$ = 1.54184 \AA); 
the diffraction patterns were scanned over a $2\thetaup$ range from 5 to 75$^{\circ}$. 
The PDF-2 Powder Diffraction Database (Joint Committee on Powder Diffraction Standards\,--\,International Centre for Diffraction Data) was used for the phase composition analysis. 

For the examination of samples using polarizing microscopy (PM) a POLAM L213-M (LOMO) microscope was employed.
Film polarizers rather than Nicol prisms are applied in the design of this instrument.

\begin{figure}[t]
	\includegraphics[width=\textwidth]{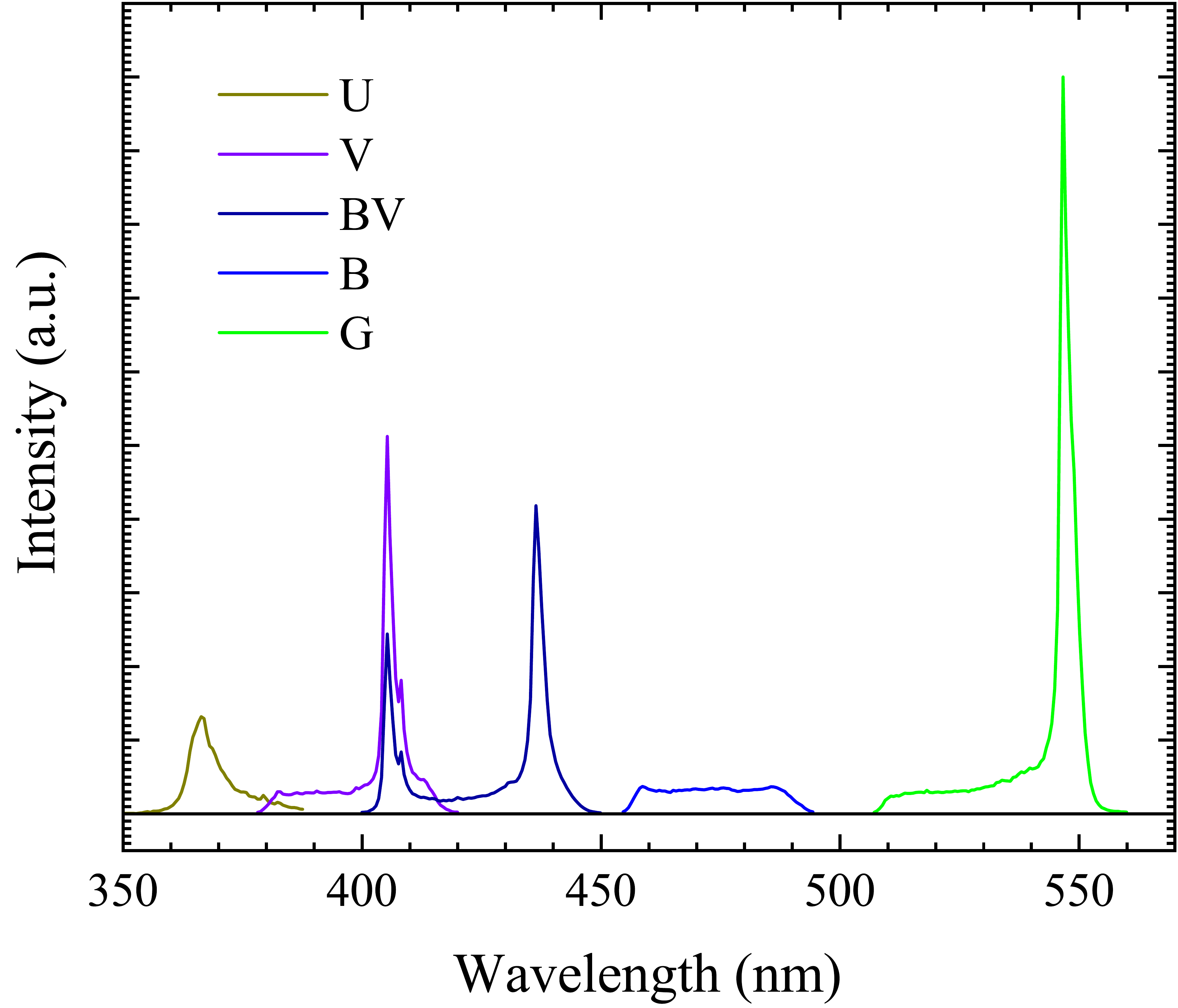}
	\caption{\label{fig:PL-EX_Spectral_Bands}%
		Mercury arc lamp spectral bands used for the excitation of PL in the luminescence microscope (the band designations---U, V, BV, B and G---correspond to those in Table~\ref{tab:PL-EX_Spectral_Bands}; 
		the input flange of the spectrometer fiber-optical cable was set coaxially with the microscope objective in place of a sample on the microscope stage during the measurements;
		technical data for the PE and receive paths of the microscope are given in Table~\ref{tab:PL-EX_Spectral_Bands}).
	}
\end{figure}

Micro photoluminescence (PL) spectra were acquired at room temperature using 
AvaSpec-2048 fiberoptic spectrometer (Avantes) connected with MIKMED-2~fluorescence stereo microscope (LOMO).
Emission bands of atomic Hg and Xe of a 100-W mercury arc lamp (HBO~100, Osram) were employed for PL exciting.
Five preinstalled microscope operation modes differing by the photoexcitation (PE) and data acquisition (DA) spectral ranges---designated as G (green), B (blue), BV (blue-violet), V (violet) and U (ultraviolet)---enabled varying the PE conditions.
Spectral data for the operation modes are given in Table~\ref{tab:PL-EX_Spectral_Bands}, and the spectral bands available for the PL excitation are plotted for each mode of the microscope in Fig.~\ref{fig:PL-EX_Spectral_Bands}.
The PL spectra of the samples studied in this article were generally obtained at the microscope magnification of $500\times$.
However, the magnification of 250$\times$ or 125$\times$ was also used sometimes.

\begin{table}[t]
	\caption{%
		Spectral data for the PE and receive paths of the PL microscope (for the spectra of the PE bands, see Fig.~\ref{fig:PL-EX_Spectral_Bands}).
		\label{tab:PL-EX_Spectral_Bands}
	}
	\begin{center}
		{\small
	\begin{tabular}{cccc}
		\hline
		Band  &PE filter  &Dichroic mirror &DA filter   \\
		designation	& transmission band  & splitting wavelength & cut-on wavelength \\
		&  (nm)              &		(nm) 			&	(nm)			\\
		\hline
		G & 510 to 560  & 575 & 590 \\
		B & 450 to 490	& 505 & 520 \\
		BV & 400 to 440	& 455 & 470 \\
		V & 380 to 420	& 430 & 450 \\
		U & 330 to 380	& 400 & 420 \\
		\hline
	\end{tabular}
}
\end{center}
\end{table}

\subsection{Specimens, sampling and sample preparation}
\label{sampling}

\subsubsection{Samples from the painting by Robert~Falk}
\label{Falk-samples}

Over 20 pigment micro samples 
covering the spectrum of hues used in the ``Birch. Spring'' painting by Robert Falk were taken under a microscope from the front and rear sides of the canvas. 
The samples were investigated using a set of analytical methods, such as SEM imaging and EDS X-ray elemental microanalysis including elemental mapping, PM and PL. 
The study showed, that the samples of the blue paint layer \#\,17 and \#\,5, which were taken at the points indicated in Fig.~\ref{fig:Falk_painting}, are of topical interest.

An examining procedure of the samples was as follows.
Before analyzing, every sample was divided into several portions;
each portion was investigated using a specific kind analysis.
SEM imaging and EDS X-ray elemental microanalysis including elemental mapping, PM and PL analyses
were routinely carried out at different portions of an original sample. 
%
 %
The originally obtained samples evidently were inhomogeneous.
However, since they were small enough to provide specimens with close properties after the division into separate portions, we consider the random variations of composition between the sample portions as insignificant.
Thus, we consider each portion of the original samples as representative, in general features reflecting the properties of the whole original sample.
This allows us to compare the properties of the original samples in whole based on data of a specific kind of analysis of their portions. 
For this reason, comparing of data obtained using specific kinds of analysis from different portions of an individual sample is also justified.

\subsubsection{Commercial Egyptian blue pigment (Kremer) used as a standard}
\label{Kremer-standard}

We used the commercial Egyptian blue pigment (Kremer \# 10060) as a standard of calcium copper tetrasilicate for the PL investigations.
Commercial art pigments are known to often be a mixture of several compounds and considerably differ in their composition depending on the manufacturer \cite{commercial_pigments_reliable} that makes it difficult to employ them as analytical standards.
This consideration made us thoroughly analyze the degree of purity of this commercial pigment to be convinced that it did not contain foreign admixtures and could be utilized as a reliable standard. 
We have investigated its elemental and phase composition using EDS, X-ray powder diffraction and polarizing microscopy.
Besides, we have explored its luminescence to determine the structure of its emission band.
PL spectra were acquired using exciting light of different PE bands of the microscope (Fig.~\ref{fig:PL-EX_Spectral_Bands}, Table~\ref{tab:PL-EX_Spectral_Bands}).

Samples of the Kremer Egyptian blue powder were not subjected to any treatment before the examinations;
they were neither washed nor etched \cite{Egyptian_Blue_New_Life}.

\section{Results and Discussion}
\label{res}

\subsection{Examination of the commercial Egyptian blue pigment}
\label{Kremer}

\subsubsection{SEM and EDS}
\label{Kremer_SEM-EDS}

\begin{figure}[t]
	\includegraphics[width=0.5\textwidth]{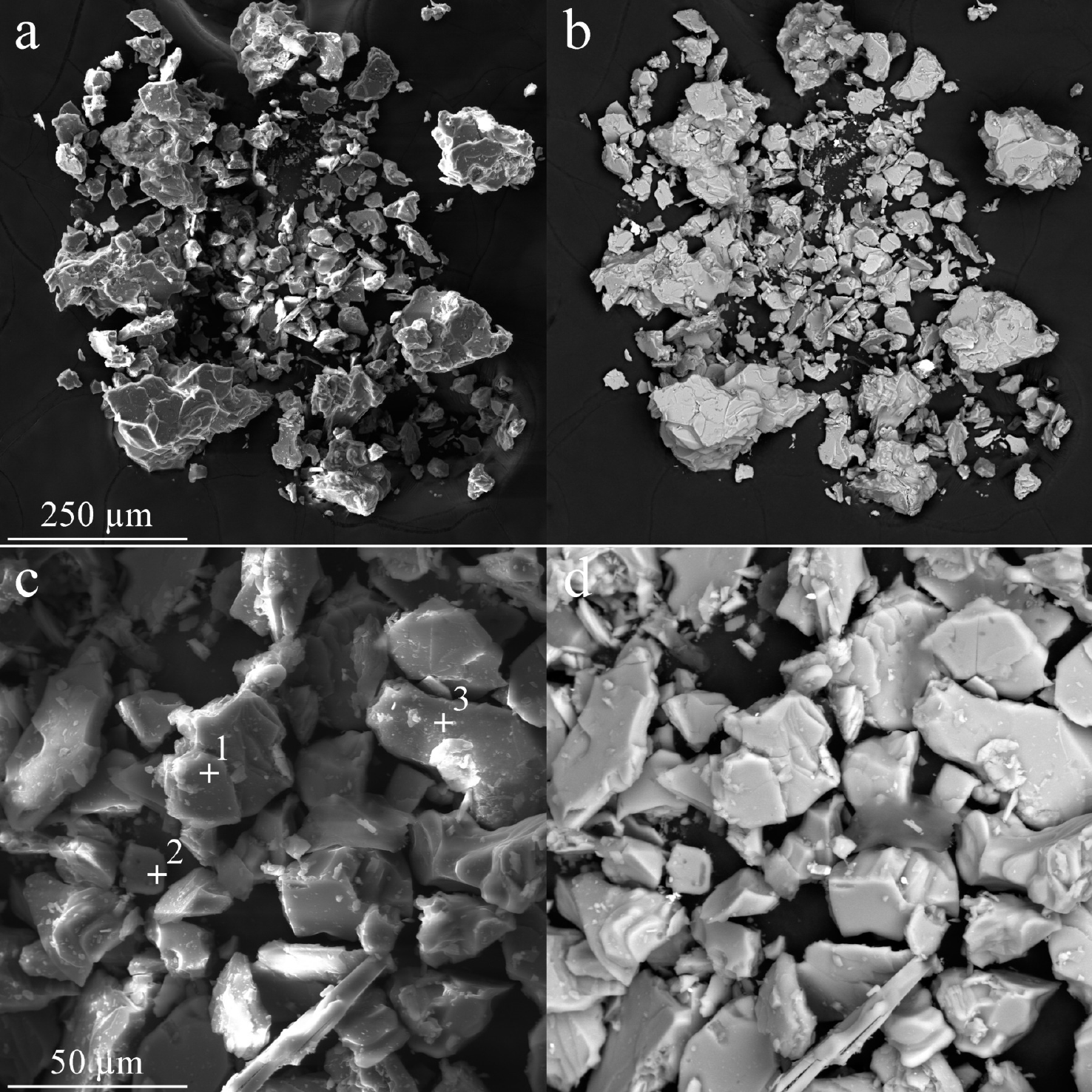}~~~
	\includegraphics[width=0.5875\textwidth]{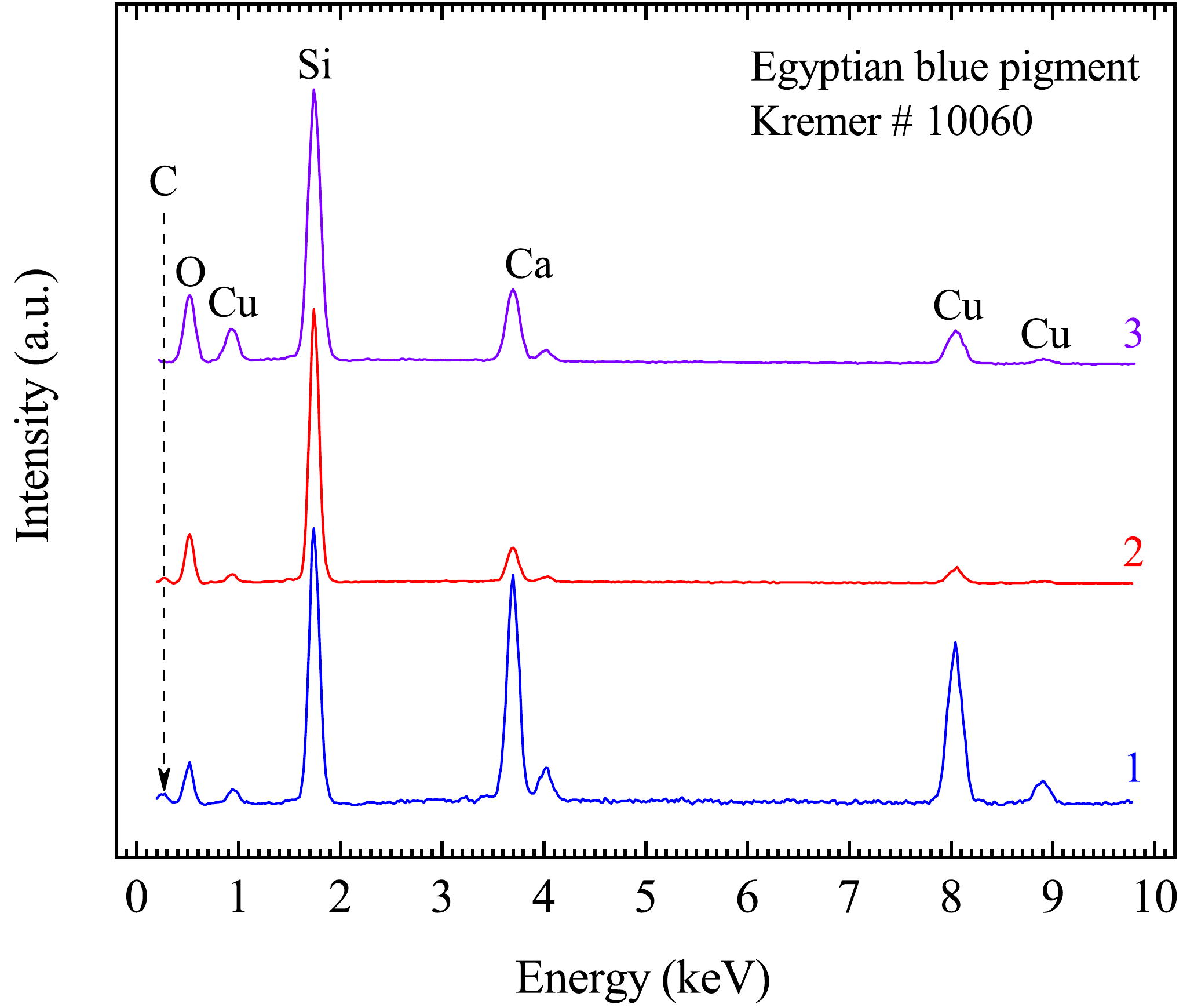}{\footnotesize \,e}{\smallskip}\\
	\includegraphics[width=0.5625\textwidth]{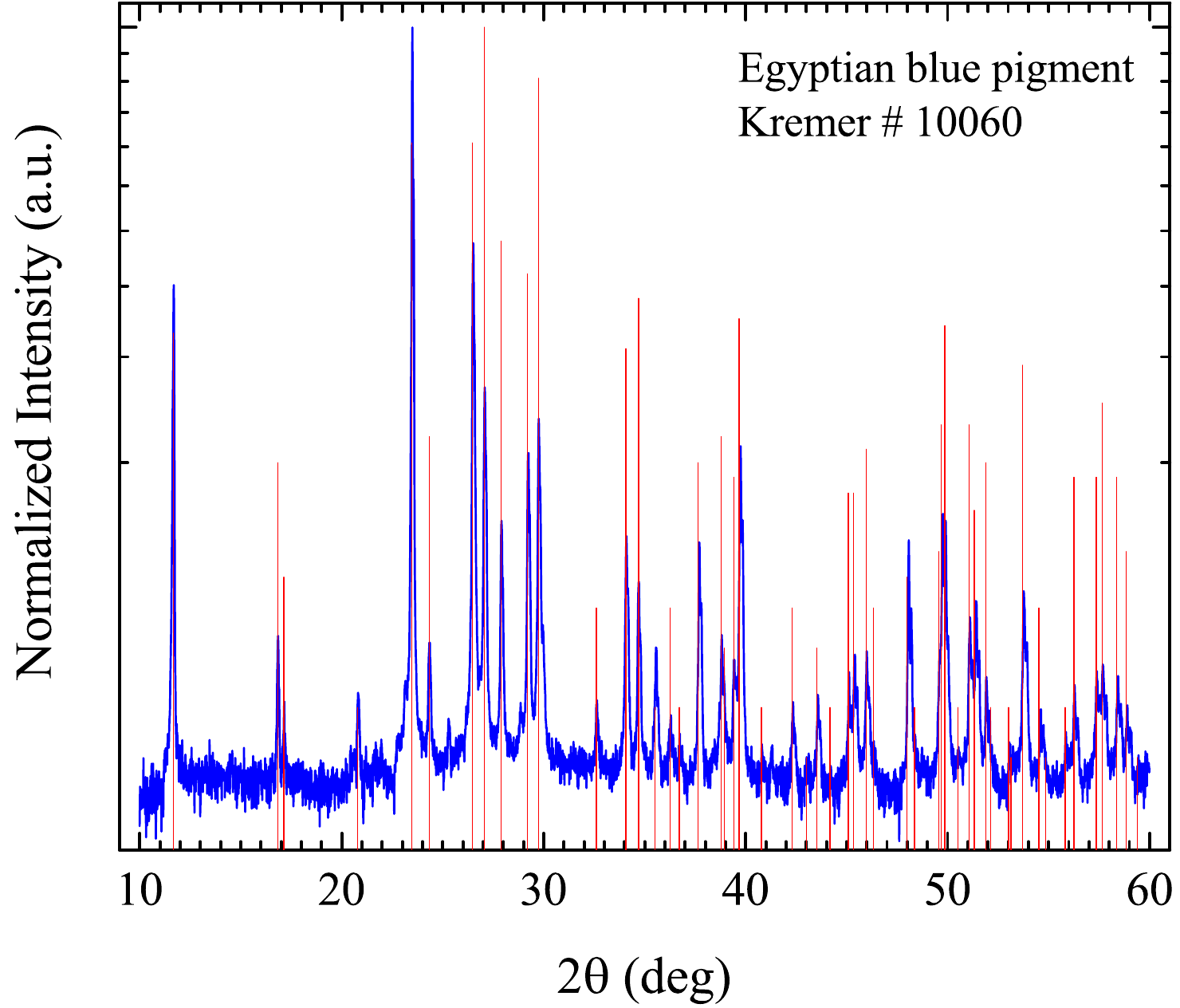}{\footnotesize \,f}
	\caption{\label{fig:Kremer_SEM_EDX_XRD}
		Pairs of SEM SE (a, c) and BSE (b, d) images of a sample of  
		the commercial Egyptian blue pigment (Kremer \#\,10060); 
		numbered crosses in the panel (c) indicate points on powder particles, at which the EDX spectra were obtained; 
		EDX spectra (e) obtained 
		at the points shown in 
		the panel (c),
		the numerals correspond to the point numbers;
		X-ray powder pattern [Cu\,K$_{\alpha}$]~(f) obtained 
		from a sample of the Kremer~\#\,10060 pigment,
		the experimental data are shown by the solid line and 
		the vertical bars present the diffraction data 
		for CaCuSi$_4$O$_{10}$
		from 
		the PDF Card 01-085-0158
		(ICSD 402012)
		\cite{Bensch:cuprorivaite}.
	}
\end{figure}

\begin{figure}[t]
	\includegraphics[width=\textwidth]{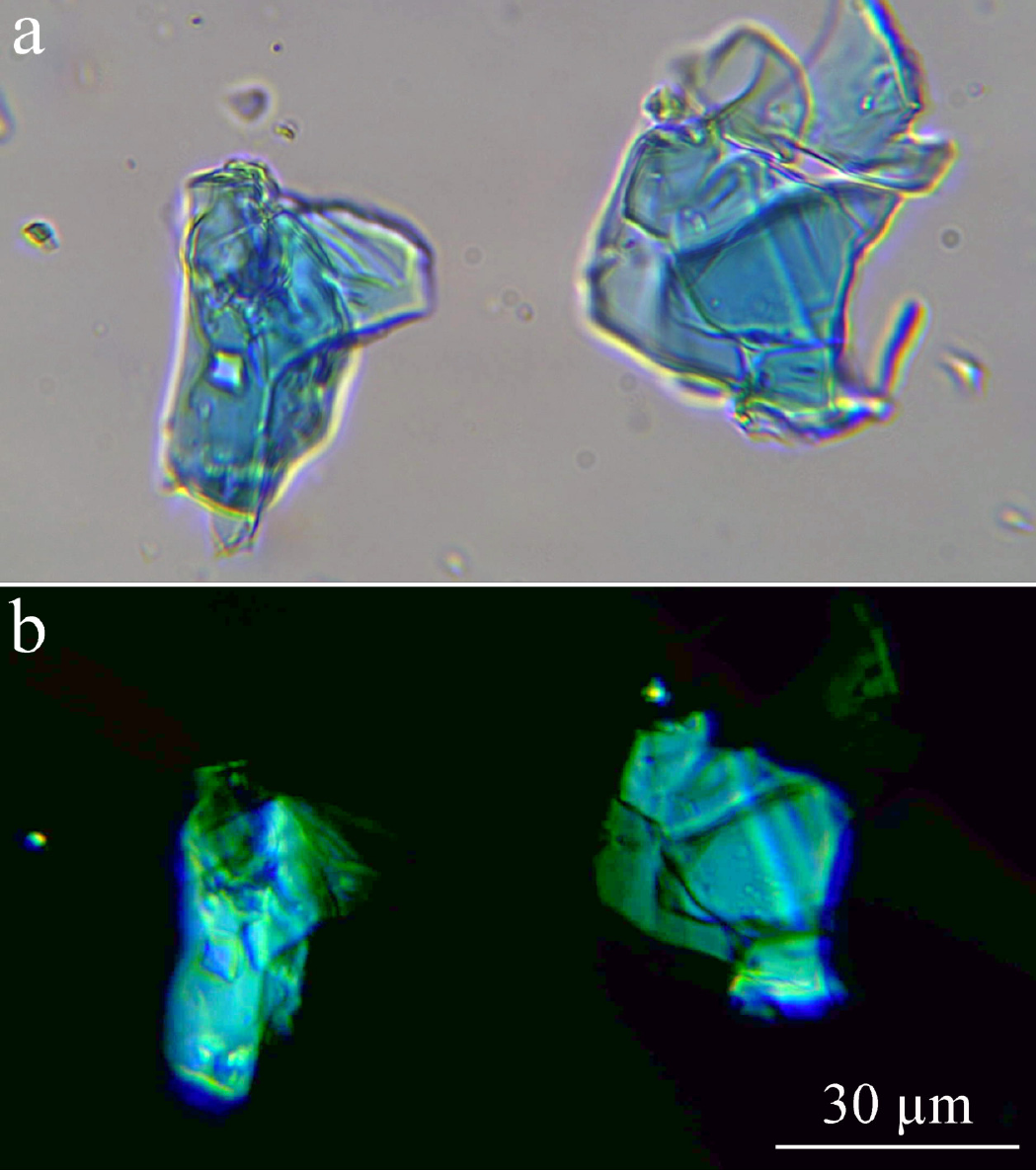}
	\caption{\label{fig:Kremer_PM}%
		PM images of powder particles of the commercial Egyptian blue pigment (Kremer \#\,10060): 
		(a)~parallel and (b)~crossed polars.
	}
\end{figure}

SEM images of samples of the investigated powder of the commercial Egyptian blue pigment are shown in Fig.~\ref{fig:Kremer_SEM_EDX_XRD}\,a--d and typical EDX spectra obtained at several points on powder particles are presented in Fig.~\ref{fig:Kremer_SEM_EDX_XRD}\,e.
SEM images obtained in backscattered electrons demonstrate the homogeneity of the powder composition.
The EDX spectra are seen to contain only peaks of oxygen, silicon, calcium and copper (the intensity of the peak of carbon likely present due to a contaminant or a residual gas of the SEM chamber atmosphere is negligible).
Although the elemental analysis of this powder was semi-quantitative, its data allow us to consider this substance as pure enough to be used as a reference sample in further analyses of paints.

\subsubsection{X-ray powder analysis}
\label{Kremer_Powder}

The X-ray powder pattern of the commercial Egyptian blue pigment shown in Fig.~\ref{fig:Kremer_SEM_EDX_XRD}\,f demonstrates that the phase of CaCuSi$_4$O$_{10}$ completely dominates its composition.
The positions of all the diffraction maxima, except for a few very weak ones that likely emerge due to parasitic reflections, coincide with the positions of those documented in the PDF Card 01-085-0158 (ICSD Entry 402012) and Ref.~\cite{Bensch:cuprorivaite}.

We conclude that the possible content of some foreign crystalline phases is negligible in this powder that allows us to employ this commercial compound as the reference one in further analyses.

\subsubsection{Polarizing microscopy}
\label{Kremer_PM}



The study of samples of the commercial Egyptian blue pigment in the plane-polarized light shows optical properties, which are characteristic to Egyptian blue \cite{Pigment_Compendium_Optical}. 
The thin bright blue translucent flat particles typical for Egyptian blue exhibit a very weak pleochroism.%
\footnote{%
	Compare, e.g., with the Egyptian blue PM images given in Ref.~\cite{Micro-EB_Arch_England}.
}
Under crossed polars, the compound displays very low order interference
colors and weak extinction.
The shape of the particles is very diverse and characterized by irregular, angular fractures (Fig.~\ref{fig:Kremer_PM}). 
Besides, a broad particle size distribution from medium (1 to 3~{\textmu}m) to very coarse ($>40$~{\textmu}m) is observed (Figs.~\ref{fig:Kremer_SEM_EDX_XRD}\,a--d and~\ref{fig:Kremer_PM}).%
\footnote{\label{fn:Feller}%
According to Feller and Bayard's particle size classification for pigments \cite{Artists_Pigments_book_PM}.%
}
%


\begin{figure}[t]
	\includegraphics[width=\textwidth]{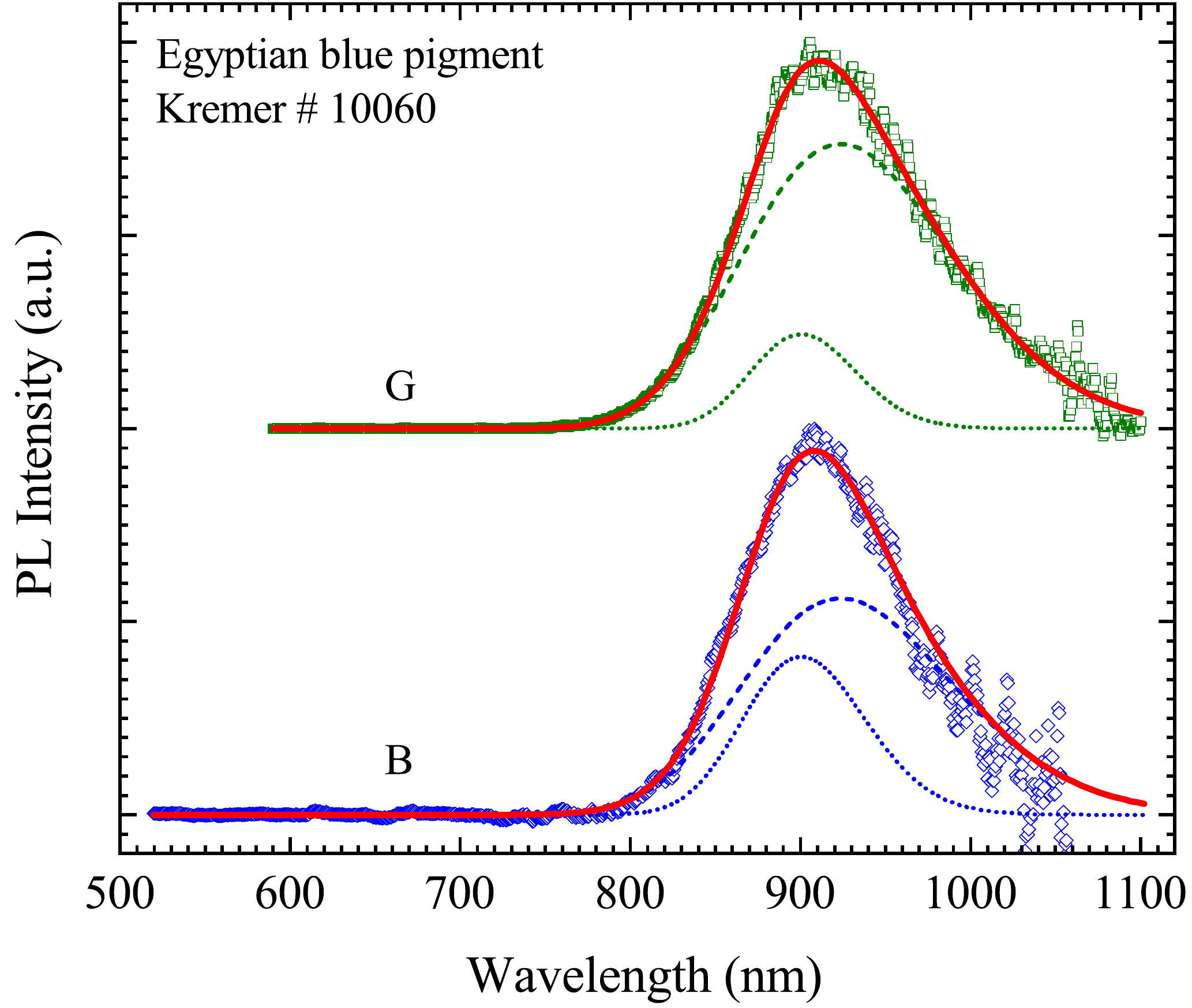}
	\caption{\label{fig:Kremer_PL}%
		PL spectra obtained from 
		a sample of  
		the commercial Egyptian blue pigment (Kremer \#\,10060);
		letters B and G designate the spectral ranges of light used for the luminescence excitation (Fig.~\ref{fig:PL-EX_Spectral_Bands}, Table~\ref{tab:PL-EX_Spectral_Bands});
		the symbols present experimental points,
		the dotted and dashed lines show the band components derived by deconvolution  
		of the experimental curves  
		using Voigt functions, 
		and 
		the solid lines are a cumulative fit peaks.
			}
\end{figure}

\subsubsection{Photoluminescence}
\label{Kremer_PL}

We have recorded PL spectra of the commercial Egyptian blue samples at a number of points of each sample and obtained similar results for them.
Spectra of PL excited using the G and B bands (Fig.~\ref{fig:PL-EX_Spectral_Bands}) nearly coincide;
as it is seen in Fig.~\ref{fig:Kremer_PL},
they consist of the only spectral band peaking at the wavelength of about 910 nm.
However, the deconvolution%
\footnote{%
	We used Voigt functions for the peak analysis.%
}
of this band for PL excited by the G and B bands demonstrates that they are composed of two Gaussian spectral bands peaking at about 900 and about 923~nm somewhat differing in the ratio of maximum intensities for different excitation bands (Fig.~\ref{fig:Kremer_PL}).%
\footnote{%
	 The ratio of maximum intensities of the components peaking at 923 and 900~nm 
	 is
	 $\sim3$ for the G band
	 and $\sim 1.5$ for the B band;
	 the peak area ratios are $\sim 5.7$ and $\sim 2.3$, respectively. %
}

PL was not detected under the sample excitation by light from the other (BV, V or U) spectral ranges.
This allows us to conclude that no other luminescent admixtures emitting visible or near infrared light are contained in the examined commercial Egyptian blue pigment that enables its utilization as a standard for the PL spectral analysis in these spectral ranges.


\subsection{Analysis of samples from the painting by Robert~Falk}
\label{Falk_Analysis}

\subsubsection{SEM and EDS}
\label{Falk_SEM-EDS}


\begin{table}[t]
	\caption{%
		Content of some chemical elements in a portion of the sample 
		obtained at 
		the point \#\,5 
		on the painting by Falk 
		(Fig.~\ref{fig:Falk_painting})
		measured using EDS analysis
		(at.\%);
		the points of analysis are shown in Fig.~\ref{fig:Falk_SEM-EDX_5}\,a;
		the corresponding EDX spectra are given in (Fig.~\ref{fig:Falk_EDS}\,a);
		the multilayered elemental composition map is presented in Fig.~\ref{fig:Falk_SEM-EDX_5}\,b.
		\label{tab:EDS_5}
	}
	\begin{tabular}{ccccccccc}
		\noalign{\smallskip}\hline
		{Chemical} & \multicolumn{7}{c}{Point number}& Mean over \\ \cline{2-8}
		element	&  51 & 52 & 53 & 54 & 55 & 56 	& 57 & the map \\ \hline
		Na		&       &       &       &       &       &       &       &  9.53	\\
		Al		&       &       &       & 18.45 &       &  4.03 &  1.59 &  9.35	\\	
		Si		& 68.82 &       &       &       &       & 42.60 & 64.83 &  7.70 \\
		Ca		& 15.51 &       &       &       &       & 12.20 & 13.61 &  2.78	\\
		Cr		&       &       & 72.62 &  3.67 &	    & 82.23	&  0.97 &  6.78	\\
		Co		&       &       &       &       &       &       &       &  1.87 \\	
		Cu		& 13.08 &       &       &       &       & 18.54 & 12.58 &  1.70 \\
		Zn		&  2.59 & 16.88 & 20.48 & 67.33 & 13.56 & 14.33 &  5.41 & 28.30	\\	
		Pb		&       & 83.12 &  6.90 & 10.55 &  4.22 & 8.30 &   1.01 & 22.40	\\									
		\hline
	\end{tabular}
\end{table}


\begin{table}[t]
	\caption{%
		Content of some chemical elements in a portion of the sample 
		obtained at 
		the point \#\,17 
		on the painting by Falk 
		(Fig.~\ref{fig:Falk_painting})
		measured using EDS analysis
		(at.\%);
		the points of analysis are shown in Fig.~\ref{fig:Falk_SEM-EDX_17}\,a;
		the corresponding EDX spectra are presented in (Fig.~\ref{fig:Falk_EDS}\,b);
		the multilayered elemental composition map is presented in Fig.~\ref{fig:Falk_SEM-EDX_17}\,b.
		\label{tab:EDS_17}
	}
	\begin{tabular}{ccccc}
		\noalign{\smallskip}\hline
		{Chemical} & \multicolumn{3}{c}{Point number}& Mean over \\ \cline{2-4}
		element	&  171   & 172 	& 173   & the map \\ \hline
		Na		&       &       &       & 12.58	\\
		Al		&       &	    &  6.86 & 16.26	\\	
		Si		& 69.45 & 68.89	& 44.87	&  9.09	\\
		Ca		& 15.23 & 16.56 & 12.63 &  1.48	\\
		Co		&       &       &       &  2.85 \\	
		Cu		& 12.84 & 14.55 & 11.18 &       \\
		Zn		& 1.22  &       & 10.08 & 25.36	\\	
		Pb		& 1.26  &       & 14.38 & 32.39	\\									
		\hline
	\end{tabular}
\end{table}

\begin{figure}[t]
\begin{minipage}[l]{1\textwidth}{\vspace{-2cm}}
	\includegraphics[width=\textwidth]{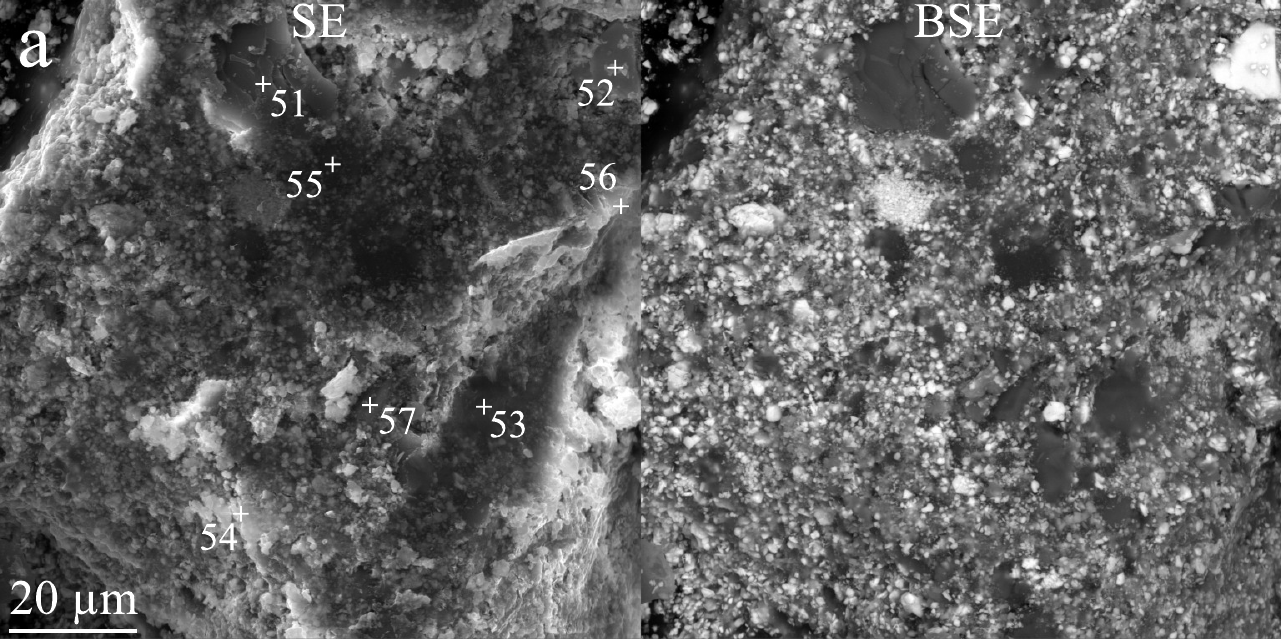}\\
	\includegraphics[width=0.25\textwidth]{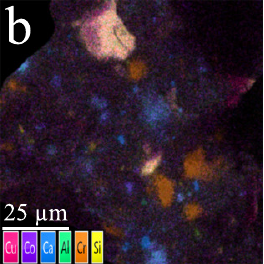}%
	\includegraphics[width=0.25\textwidth]{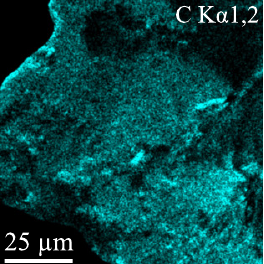}%
	\includegraphics[width=0.25\textwidth]{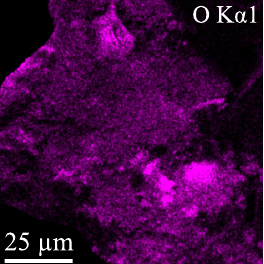}%
	\includegraphics[width=0.25\textwidth]{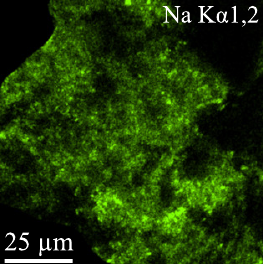}\\%
	\includegraphics[width=0.25\textwidth]{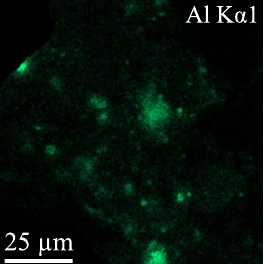}%
	\includegraphics[width=0.25\textwidth]{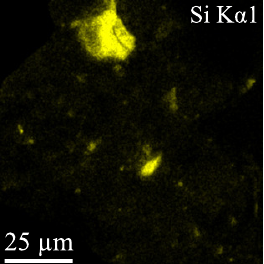}%
	\includegraphics[width=0.25\textwidth]{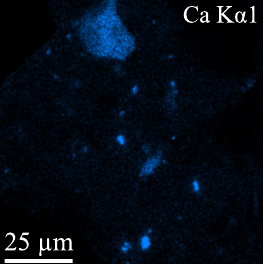}%
	\includegraphics[width=0.25\textwidth]{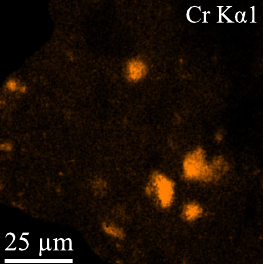}\\%
	\includegraphics[width=0.25\textwidth]{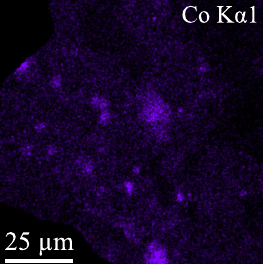}%
	\includegraphics[width=0.25\textwidth]{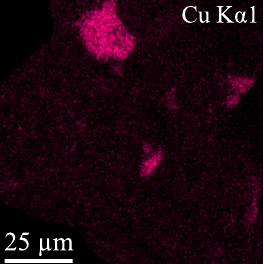}%
	\includegraphics[width=0.25\textwidth]{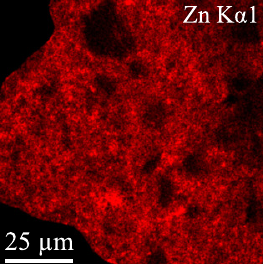}%
	\includegraphics[width=0.25\textwidth]{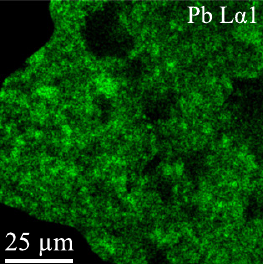}
\end{minipage}
	\caption{\label{fig:Falk_SEM-EDX_5}%
		A pair of SEM SE and BSE images (a) and elemental composition maps (b) 
		of a specimen taken from the paint layer samples from the point \#\,5 of the painting by Falk 
		(Fig.~\ref{fig:Falk_painting});
		numbered crosses in the panel (a) indicate points on the specimen, at which the EDX spectra were recorded (Fig.~\ref{fig:Falk_EDS}\,a);
		single-colored maps represent the distributions of signals of the X-ray bands of a number of chemical elements  
		(the bands are given in the upper right corner of each map);
		a multicolored map is a multilayered one composed of the X-ray band maps of several elements (the elements are shown in the lower left corner of this map).
	}
\end{figure}

\begin{figure}[t]
\begin{minipage}[l]{1\textwidth}{\vspace{-2cm}}
	\includegraphics[width=\textwidth]{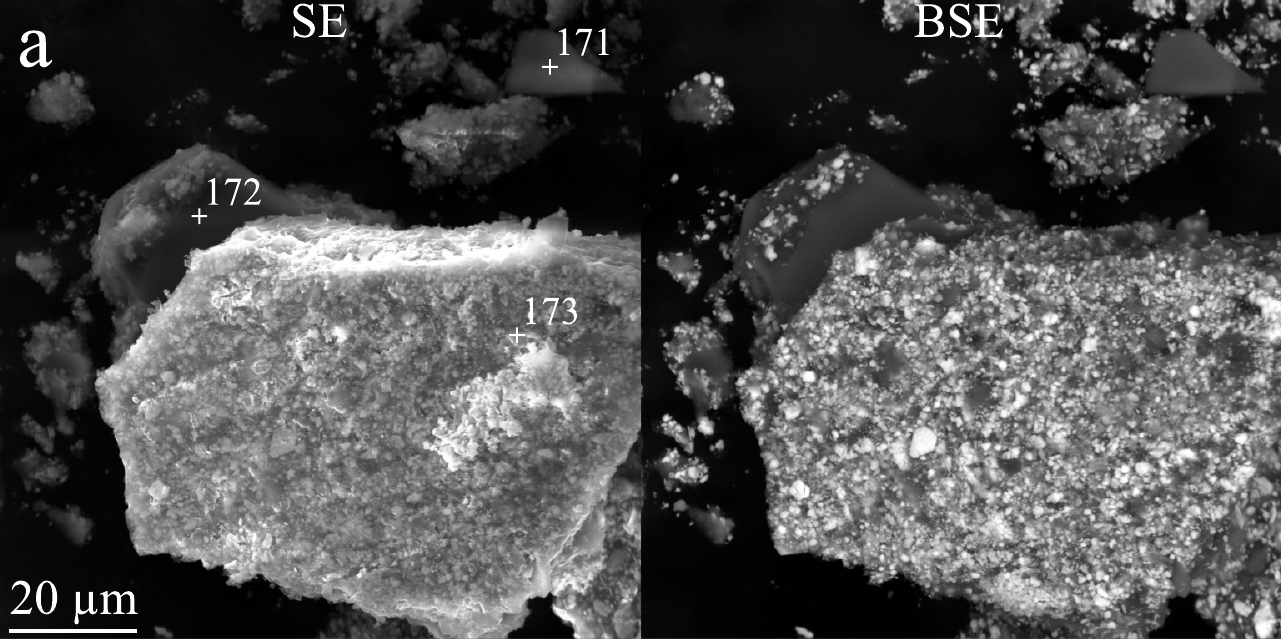}\\
	\includegraphics[width=0.25\textwidth]{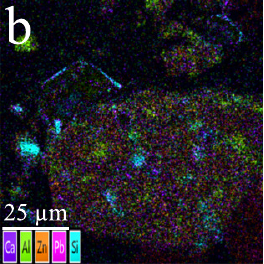}%
	\includegraphics[width=0.25\textwidth]{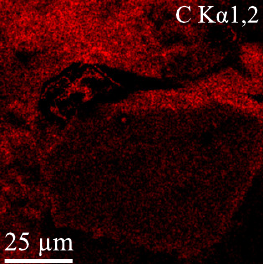}%
	\includegraphics[width=0.25\textwidth]{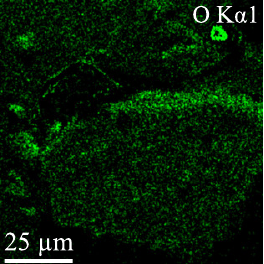}%
	\includegraphics[width=0.25\textwidth]{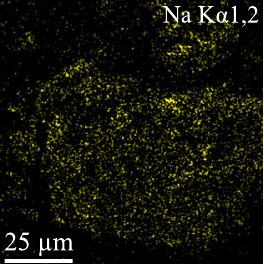}\\%
	\includegraphics[width=0.25\textwidth]{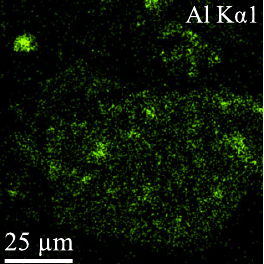}%
	\includegraphics[width=0.25\textwidth]{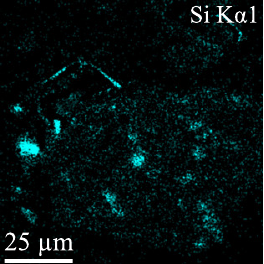}%
	\includegraphics[width=0.25\textwidth]{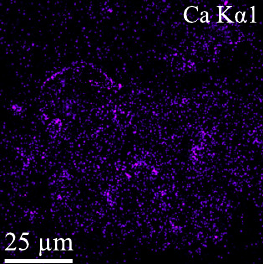}%
	\includegraphics[width=0.25\textwidth]{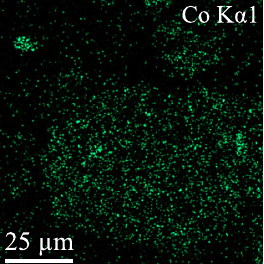}\\%
	\includegraphics[width=0.25\textwidth]{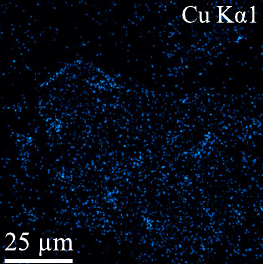}%
	\includegraphics[width=0.25\textwidth]{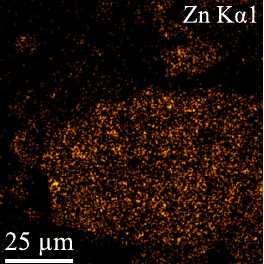}%
	\includegraphics[width=0.25\textwidth]{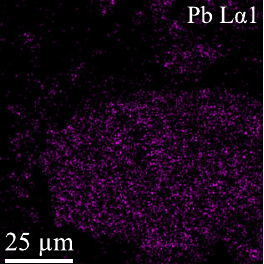}
\end{minipage}
	\caption{\label{fig:Falk_SEM-EDX_17}%
		A pair of SEM SE and BSE images (a) and elemental composition maps (b) 
		of a specimen taken from the paint layer samples from the point \#\,17 of the painting by Falk 
		(Fig.~\ref{fig:Falk_painting});
		numbered crosses in the panel (a) indicate points on the specimen, at which the EDX spectra were recorded (Fig.~\ref{fig:Falk_EDS}\,b);
		single-colored maps represent the distributions of signals of the X-ray bands of a number of chemical elements  
		(the bands are given in the upper right corner of each map);
		a multicolored map is a multilayered one composed of the X-ray band maps of several elements (the elements are shown in the lower left corner of this map).
	}
\end{figure}

\begin{figure}[t]
	\includegraphics[width=0.5\textwidth]{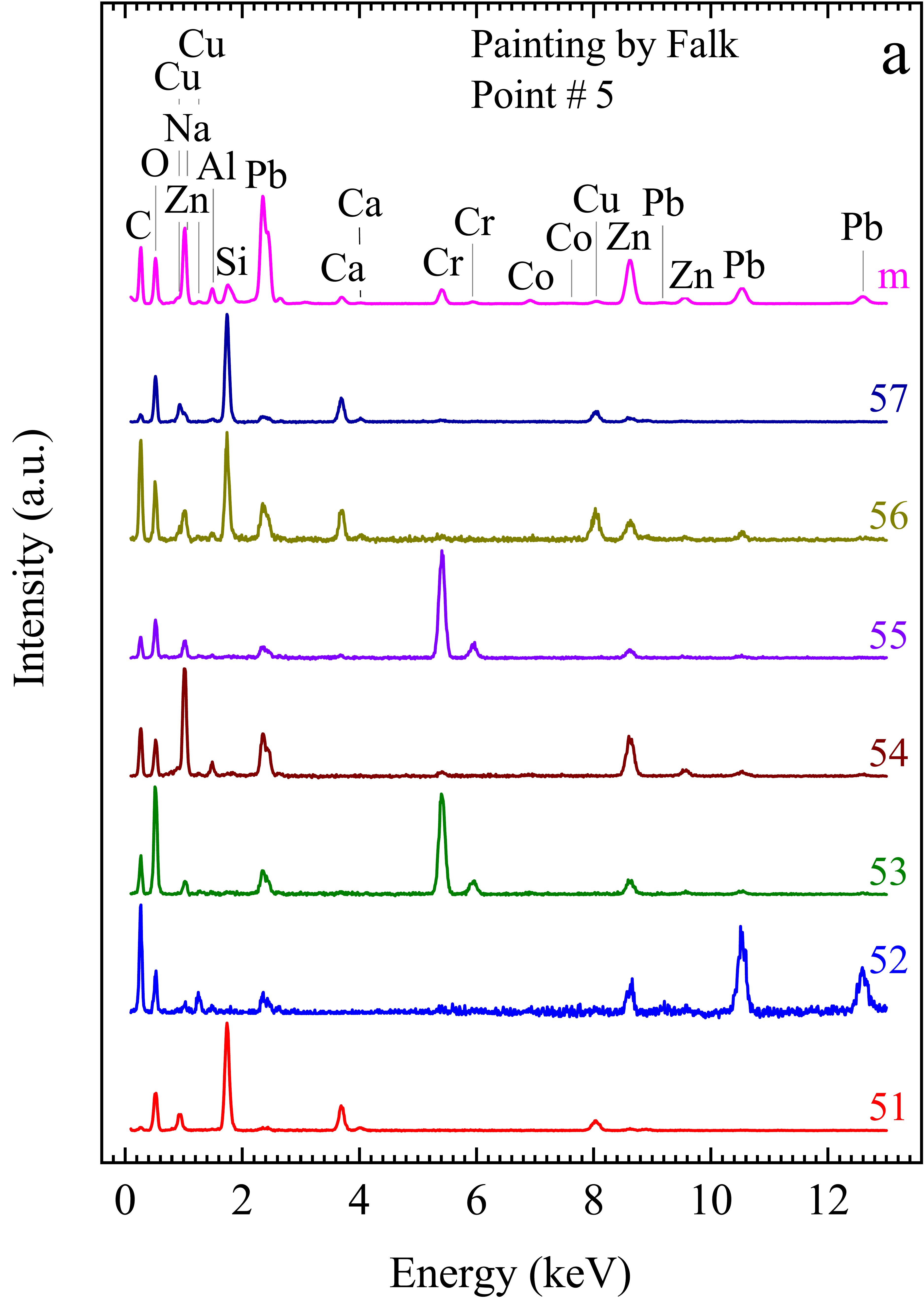}~~~
	\includegraphics[width=0.5\textwidth]{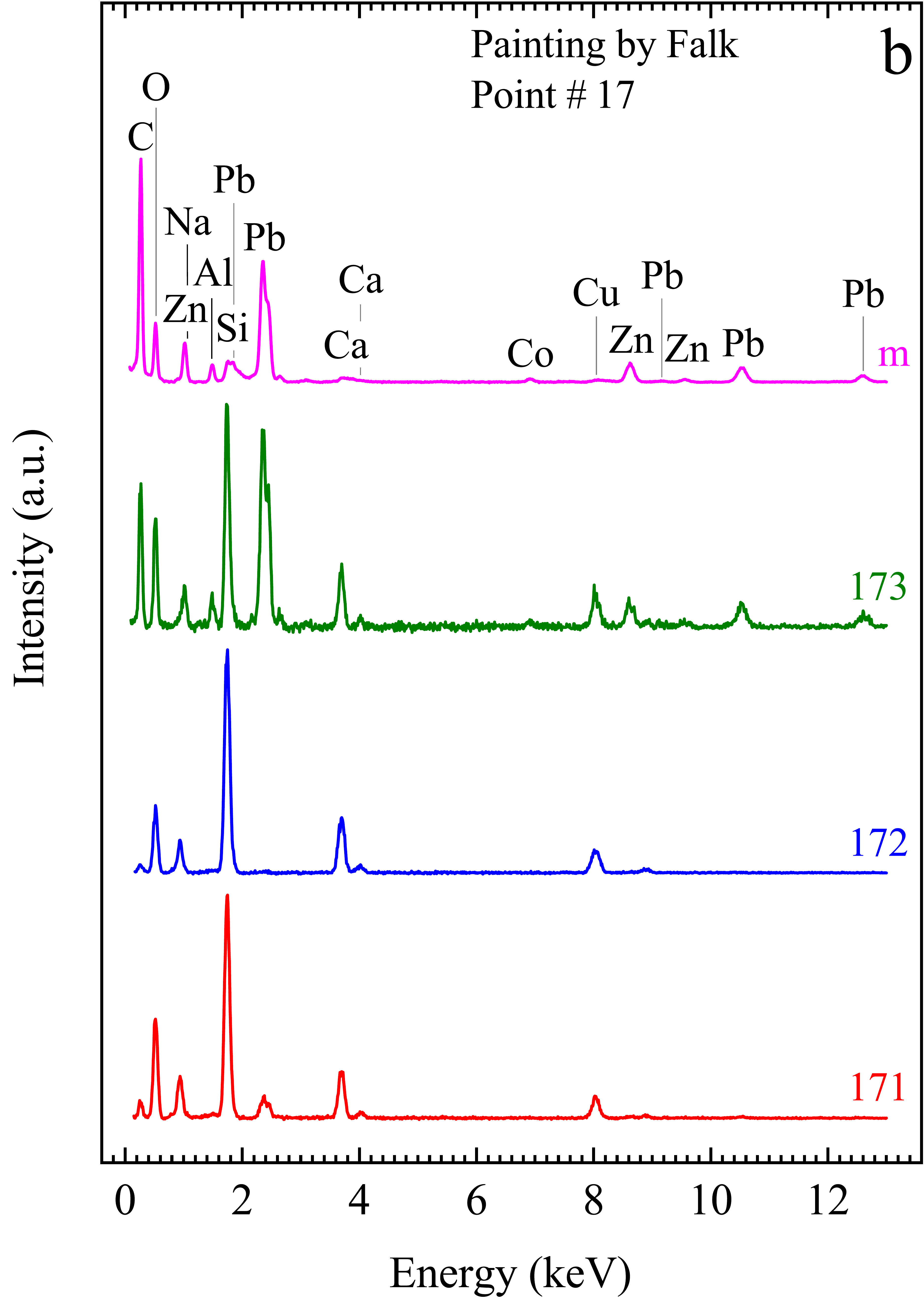}
	\caption{\label{fig:Falk_EDS}%
		EDX spectra 
		obtained at separate points on specimens taken from the paint layer samples from the points \#\,5 (a) and \#\,17 (b) of the painting by Falk 
		(Fig.~\ref{fig:Falk_painting})
		and by averaging X-ray spectra over the elemental composition map of each specimen;
		numerals at the curves correspond to the point numbers (Figs.~\ref{fig:Falk_SEM-EDX_5}\,a and~\ref{fig:Falk_SEM-EDX_17}\,a), the letter `m' designates the mean spectrum of the corresponding elemental composition map (Figs.~\ref{fig:Falk_SEM-EDX_5}\,b and~\ref{fig:Falk_SEM-EDX_17}\,b).
	}
\end{figure}

Specimens taken from the samples obtained at the points \#\,5 and \#\,17 on the painting by Falk were examined using SEM operated in both SE and BSE modes.

Fig.~\ref{fig:Falk_SEM-EDX_5}\,a presents a pair of such SEM images typical for the point \#\,5.
As it is seen from the BSE image, the paint layer specimen is composed of particles of substances differing in their composition and sizes.

Maps of spatial distribution of X-ray bands corresponding to a number of chemical elements recorded using the SEM EDX spectrometer at the same areas on the specimen are shown in Fig.~\ref{fig:Falk_SEM-EDX_5}\,b.
The signal of the silicon K$_{\alpha_1}$ band is seen to spatially correlate with the signals of the calcium K$_{\alpha_1}$, copper K$_{\alpha_1}$ and oxygen K$_{\alpha_1}$ bands in these maps.
At the same time, it anti-correlates with the signals of the carbon K$_{\alpha_{1,2}}$, sodium K$_{\alpha_{1,2}}$ and zinc K$_{\alpha_1}$ and lead L$_{\alpha_1}$ bands.
No spatial correlation of this band was registered with the aluminum K$_{\alpha_1}$, chromium K$_{\alpha_1}$ and cobalt K$_{\alpha_1}$ bands.

However, some domains of the map of the oxygen K$_{\alpha_1}$ band are seen to correlate with the map of the chromium K$_{\alpha_1}$ band.
Besides, O\,K$_{\alpha_1}$ signal is seen to be distributed quite uniformly across the map as well as that of C\,K$_{\alpha_{1,2}}$, Zn\,K$_{\alpha_1}$ and Pb L$_{\alpha_1}$ and all these signals correlate well with one another.
Additionally, some bright features of the O\,K$_{\alpha_1}$ map coincide with some bright features of the C\,K$_{\alpha_{1,2}}$ map.

The signals of Al\,K$_{\alpha_1}$ and Co\,K$_{\alpha_1}$ also strongly correlate with one another.
The Na\,K$_{\alpha_{1,2}}$ band is distributed quite homogeneously across the map with numerous tiny bright spots that are also spread rather uniformly in the map. 
The uniformity of the Na\,K$_{\alpha_{1,2}}$ distribution and the smallness of sizes of its bright features make it difficult to detect any correlation between this map and the rest ones.
Nevertheless, some bright points in the Si\,K$_{\alpha_1}$ and Na\,K$_{\alpha_{1,2}}$ maps are seen to coincide (e.g., a couple of small bright spots at the center of the left edge of the Si\,K$_{\alpha_1}$ map and a pair of similar spots at the same place of the Na\,K$_{\alpha_{1,2}}$ map).

The multilayered map of the spatial distribution of the Cu, Co, Ca, Al, Cr and Si K$_{\alpha_1}$ X-ray bands clearly demonstrates the presence of compounds, which contain a combination of Co and Al and a combination Cu, Ca and Si, gathered in particles in this specimen.
Besides, separate areas containing Cr are also seen in the map.

A pair of SEM SE and BSE images typical for the point \#\,17 is shown in Fig.~\ref{fig:Falk_SEM-EDX_17}.
In general features, the composition of this paint layer specimen (see the BSE image in Fig.~\ref{fig:Falk_SEM-EDX_17}\,a) looks similar to the composition of the specimen taken at the point \#\,5 (the BSE image in Fig.~\ref{fig:Falk_SEM-EDX_5}\,a) with the exception of the sizes of most particles, which look somewhat smaller than those in the sample taken in the point \#\,5.

Maps of the X-ray bands distribution (Fig.~\ref{fig:Falk_SEM-EDX_17}\,b) also demonstrate the same regularities as the maps of the sample from the point \#\,5.
The maps of the Si\,K$_{\alpha_1}$, Ca\,K$_{\alpha_1}$ and Cu\,K$_{\alpha_1}$ lines demonstrate coinciding features.
Some features in the map of O\,K$_{\alpha_1}$ coincide on the specimen surface with the analogous features in the maps of Al\,K$_{\alpha_1}$, Co\,K$_{\alpha_1}$, Si\,K$_{\alpha_1}$, Ca\,K$_{\alpha_1}$ and Cu\,K$_{\alpha_1}$.
Distributions of Zn\,K$_{\alpha_1}$ and Pb\,L$_{\alpha_1}$ are quite similar; they are distributed quite uniformly.

The multilayered map of the specimen from the point \#\,17 shows that, judging from the spatial distribution of the Ca K$_{\alpha_1}$, Al K$_{\alpha_1}$, Si K$_{\alpha_1}$ and Pb\,L$_{\alpha_1}$ X-ray bands, grains containing Si and particles of a substance containing Al are included in a quite uniformly distributed mixture of fine grains of compounds containing Pb or Zn.

EDX spectra recorded at the points, indicated in Figs.~\ref{fig:Falk_SEM-EDX_5}\,a and~\ref{fig:Falk_SEM-EDX_17}\,a, are shown in the panels (a) and (b) of Fig.~\ref{fig:Falk_EDS} for the points \#\,5 and \#\,17, respectively.
They evidence the presence of C, O, Na, Al, Si, Ca, Cr, Co, Cu, Zn and Pb in detectable quantities in the point \#\,5 and the same set of elements except for Cr in the point \#\,17 on the painting by Falk. 
Semi-quantitative estimates of their content in the samples from both points are given in Tables~\ref{tab:EDS_5} and~\ref{tab:EDS_17}.
It is seen from the spectra and tables that different compounds are detected in different points of the elemental analysis.

They can be easily identified if elemental mapping and EDS analysis are combined with the polarizing optical microscopy.

\subsubsection{Polarizing microscopy}
\label{Falk_PM}

\begin{figure}[t]
	\includegraphics[width=0.45\textwidth]{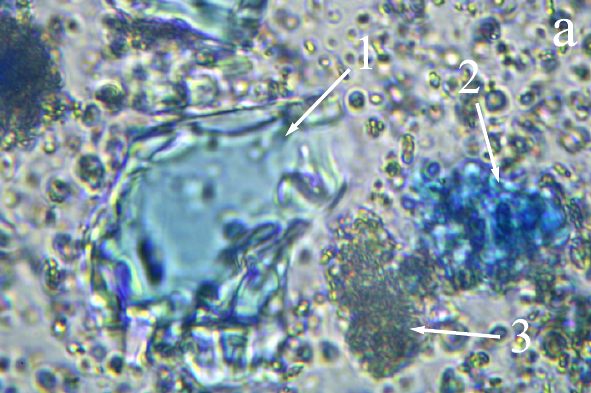}\,%
	\includegraphics[width=0.45\textwidth]{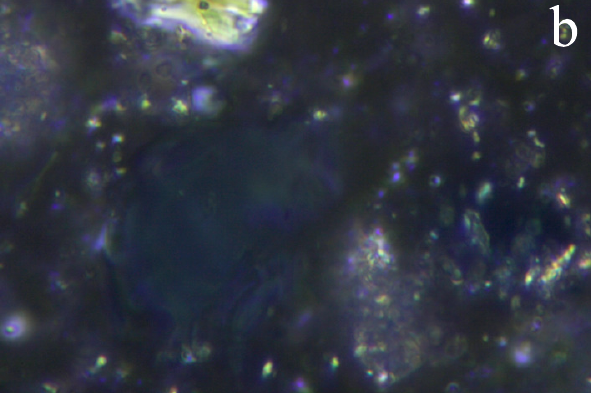}\\
	\includegraphics[width=0.45\textwidth]{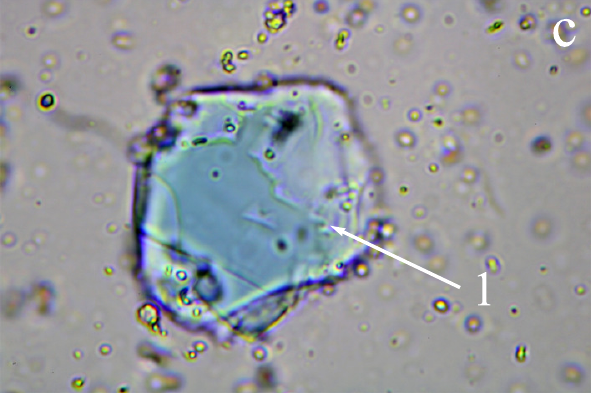}\,%
	\includegraphics[width=0.45\textwidth]{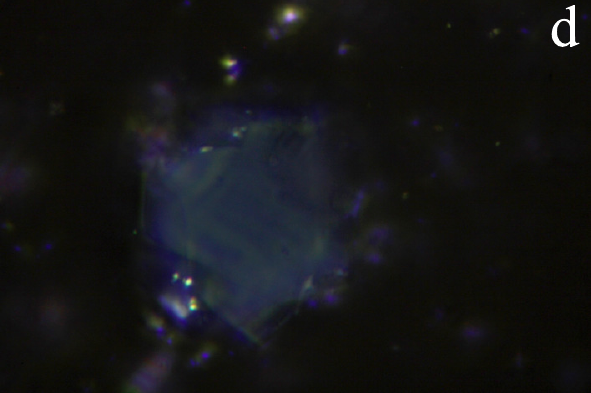}
	\includegraphics[width=0.45\textwidth]{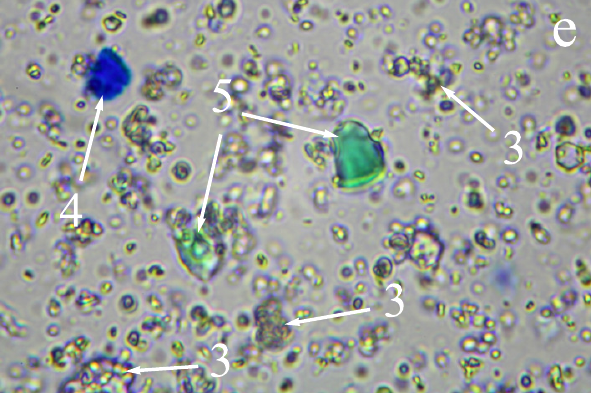}\,%
	\includegraphics[width=0.45\textwidth]{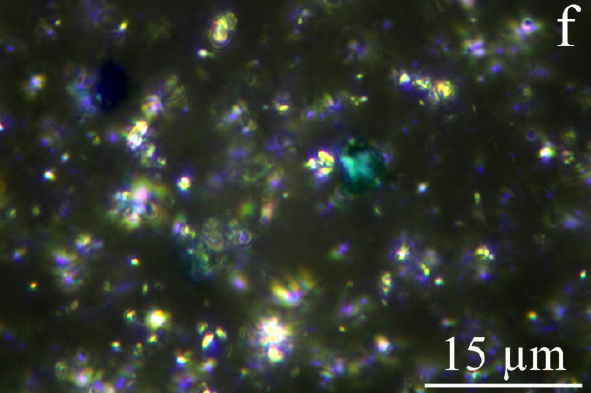}
	\caption{\label{fig:Falk_PM_5-17}%
		Complementary pairs	of  
		PM images 
		[(a) and (b), (c) and (d), (e) and (f)]
		of portions of the paint layer samples obtained at the points \#\,17 [(a), (b)] and \#\,5 [(c) to (f)] of the painting by Falk 
		(see Fig.~\ref{fig:Falk_painting}) 
		using
		 parallel [(a), (c), (e)] and~crossed [(b), (d), (f)] polars;
		 the numbered arrows indicate examples of particles of 
		 Egyptian blue (1),
		 cobalt blue (2), 
		 lead and zinc whites (3),
		 artificial ultramarine (4) 
		 and 
		 viridian (5).
	}
\end{figure}


Preliminary studies of the samples \#\,5 and \#\,17 by PM microscopy show the presence of blue crystals, which optical properties correspond to Egyptian blue (see Section~\ref{Kremer_PM}). 
However, in contrast to the commercial pigment, its color is much less intense (Fig.~\ref{fig:Falk_PM_5-17}), likely due to thinner flakes. 
The particle sizes vary from large (3 to 10~{\textmu}m) to coarse (10 to 40~{\textmu}m) (Fig.~\ref{fig:Falk_SEM-EDX_5} and~\ref{fig:Falk_SEM-EDX_17}).\cref{fn:Feller} 
In addition, zinc white (ZnO), lead white (2PbCO$_3${\textperiodcentered}Pb(OH)$_2$ + PbCO$_3$), synthetic ultramarine (Na$_7$Al$_6$Si$_6$O$_{24}$S$_{3}$), cobalt blue (CoO{\textperiodcentered}Al$_2$O$_3$) were identified in both samples, and viridian (Cr$_2$O$_3${\textperiodcentered}2H$_2$O) in sample \#\,5.%
\footnote{%
The composition was determined based on both PM and the data of SEM-EDS (Figs.~\ref{fig:Falk_SEM-EDX_5}, \ref{fig:Falk_SEM-EDX_17} and~\ref{fig:Falk_EDS}; Tables~\ref{tab:EDS_5} and~\ref{tab:EDS_17}).
} 
The ratio of these pigments in each sample is diverse and regulates the hue of the blue paint layer.

\subsubsection{Photoluminescence}
\label{Falk_PL}

\begin{figure}[t]
	\includegraphics[width=0.5\textwidth]{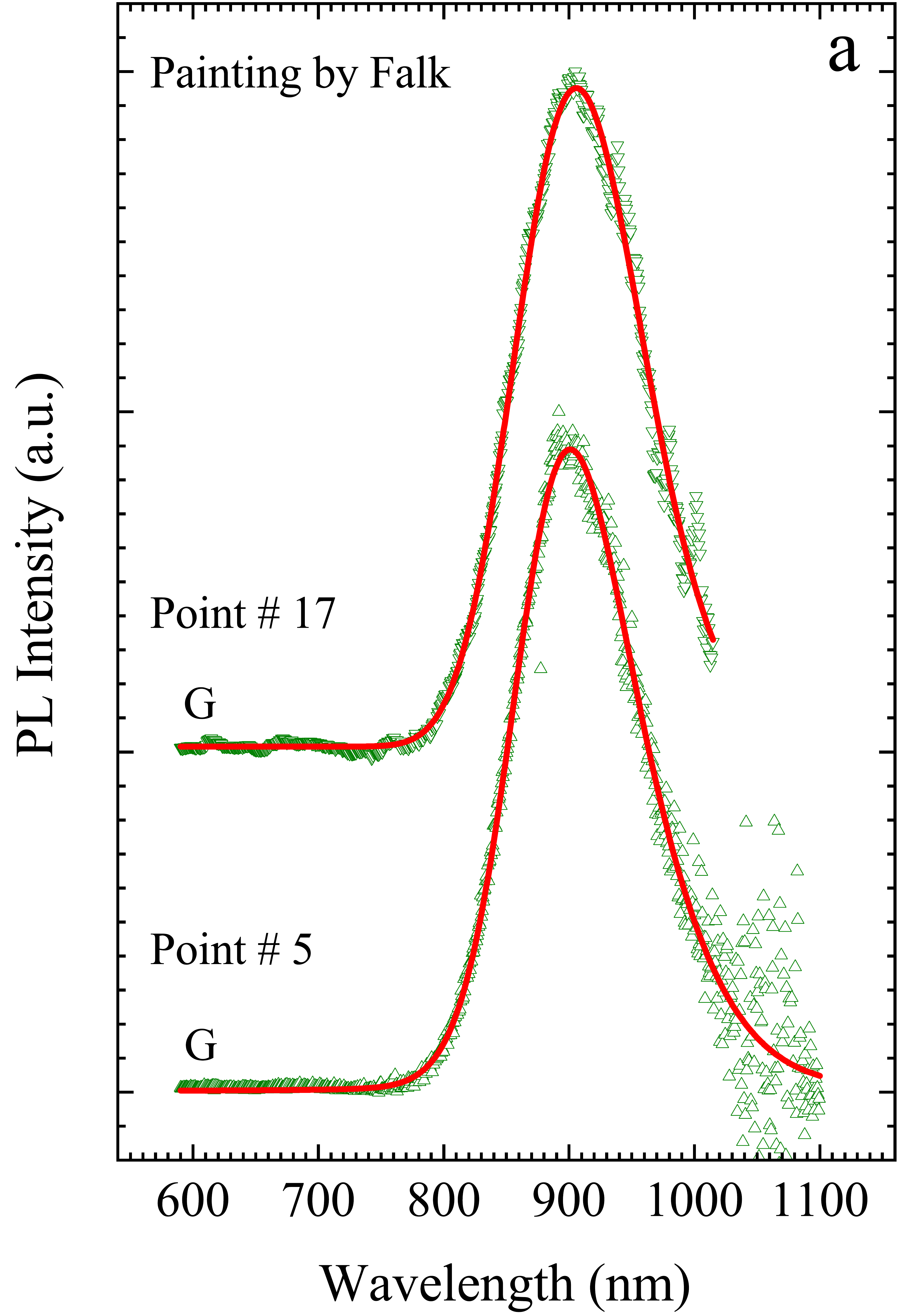}~~~
	\includegraphics[width=0.5\textwidth]{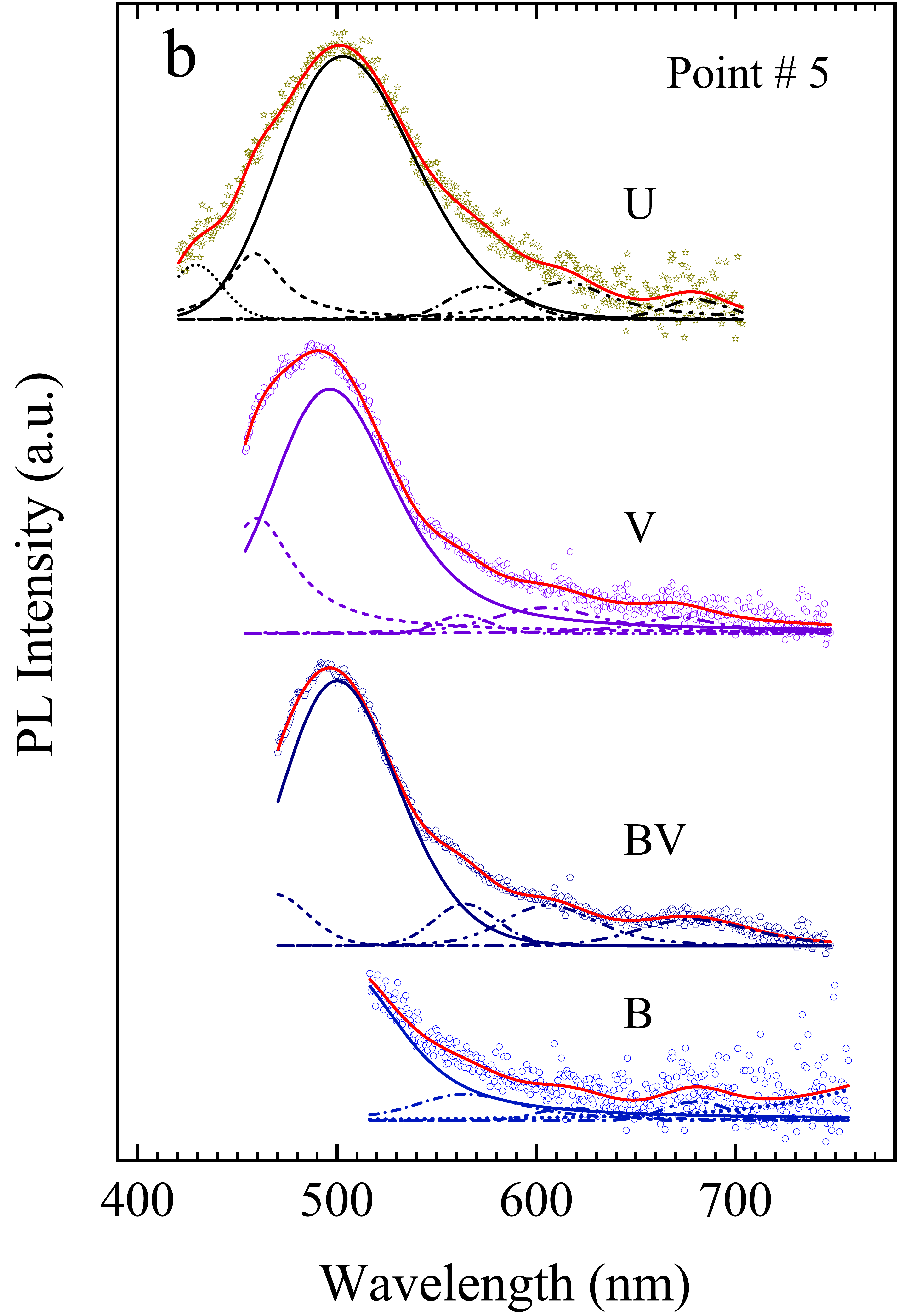}
	\caption{\label{fig:Falk_PL}%
		PL spectra 
		recorded at portions of the paint layer samples obtained at the points \#\,5 and \#\,17 of the painting by Falk
		(Fig.~\ref{fig:Falk_painting});
		designations G, B, BV, V and U indicate the spectral ranges of light used for the luminescence excitation (Fig.~\ref{fig:PL-EX_Spectral_Bands}, Table~\ref{tab:PL-EX_Spectral_Bands}).
	}
\end{figure}

PL spectra recorded from different portions of the paint layer samples obtained at the points \#\,5 and \#\,17 of the painting by Falk
(Fig.~\ref{fig:Falk_painting}) demonstrate similar sets of features specific for each PE band (Fig.~\ref{fig:Falk_PL}).
The spectra obtained under the G-band PE (Fig.~\ref{fig:PL-EX_Spectral_Bands}, 
Table~\ref{tab:PL-EX_Spectral_Bands}) in the range from 590 to 1100~nm demonstrate a single band, peaked around 906~nm, which coincides with the band of the reference sample (Fig.~\ref{fig:Kremer_PL}) and for this reason is unequivocally attributed to Egyptian blue (Fig.~\ref{fig:Falk_PL}\,a).

PL spectra obtained using the other PE bands are presented in Fig.~\ref{fig:Falk_PL}\,b.
For each PE band, the spectra were recorded in the specific spectral range presented in Table~\ref{tab:PL-EX_Spectral_Bands}.
All the spectra consist of a broad spectral band peaked at $\sim500$~nm composed of a predominating strong line and five weak lines on both sides of it (the spectra excited using the B-band show only the long-wavelength wing of this strong line). 
The strongest band is peaked at $\sim500$~nm and the weak ones are peaked around 430, 460, 560--570, 610--620 and 680 nm (the parameters of the weak lines obtained from the deconvolution of the initial bands are determined with rather high degree of ambiguity).

Remark that only the spectra for the point \#\,5 are presented in Fig.~\ref{fig:Falk_PL}\,b.
The spectra obtained for the point \#\,17 are similar to them but much more noisy because of somewhat lower intensity of PL excited using all the PE bands except for the band~G.
However, they demonstrate the same set of spectral lines as the spectra recorded at the point \#\,5 with a single predominating line at $\sim500$~nm and five much weaker ones on both sides of it.

Note also that the B-band-excited PL of the samples from both points, like in the case of the reference sample of Egyptian blue, also demonstrates a spectral band at $\sim 900$ nm, but, due to much lower intensity, the recorded signal of this PL band is too noisy and for this reason is not shown in the spectrum obtained from the sample taken from the Falk's painting (Fig.~\ref{fig:Falk_PL}\,b).

\subsection{Brief discussion}
\label{disc}

\subsubsection{On the paint layer composition and the nature of its luminescence}
\label{disc:PL}

The above analyses allow us to conclude that both of the samples from the painting by Falk contain close sets of pigments.
The sample taken from the point \#\,17 contains
{%
lead white 
(2PbCO$_3${\textperiodcentered}Pb(OH)$_2$\,+\,PbCO$_3$)
and 
zinc white (ZnO),  
artificial ultramarine 
(Na$_{8-10}$Al$_6$Si$_6$O$_{24}$S$_{3}$), 
cobalt blue 
(CoO{\textperiodcentered}Al$_2$O$_3$),
and
Egyptian blue 
(CaCuSi$_4$O$_{10}$).%
}

The composition of the sample taken from the point \#\,5 is mainly similar to that of the sample from the point \#\,17.
In addition to the pigments found in the latter sample, this one contains also 
viridian (Cr$_2$O$_3${\textperiodcentered}2H$_2$O).
Besides, it contains somewhat less amount of Egyptian blue.
For this reason, paint in the point\,\#\,5 has somewhat greener tint than in the point\,\#\,17.

PL spectral analysis unambiguously proves the presence of Egyptian blue in the paint layer of the picture by Falk since the PL spectra obtained from the samples taken at both points (Fig.~\ref{fig:Falk_PL}\,a) demonstrate the intense emission at the characteristic PL band peaking at about 910~nm. 

It should be noticed, however, that some earlier articles reported the position of the PL band characteristic to Egyptian blue nearly coinciding with the peak position presented in this work (e.g., $\lambda_{\rm max} \approx 910$~nm at $\lambda_{\rm PE} = 637$~nm \cite{Egyptian_Blue_PL}, $\lambda_{\rm max} \approx 909$~nm at $\lambda_{\rm PE} = 625$~nm \cite{Egyptian_Blue_New_Life}). 
Others, on the contrary, reported a somewhat displaced PL band for this compound (e.g., $\lambda_{\rm max} \approx 950$~nm at  $\lambda_{\rm PE} = 632.8$~nm \cite{PL_EB_HB_HP}, $\lambda_{\rm max} \approx 932$~nm \cite{EB_NIR_emission_upconversion}).
Nevertheless, all those articles assigned these PL bands to  the $^2{\rm B_{2g}}\rightarrow{\rm ^2B_{1g}}$ transition in Cu$^{2+}$ ions.%
\footnote{%
	The calcium copper silicate structure and a diagram of Cu$^{2+}$ energy levels responsible for the light emission in their connection with the crystal structure are discussed, e.g., in Refs.~\cite{EB_NIR_emission_upconversion,PL_EB_HB_HP,Egyptian_Blue_New_Life,Warner_EB_Synthesis-2011}.%
}

Moreover, the authors of Ref.~\cite{MCuSi4O10_IR_phosphors}, when investigated Ca, Ba and Sr copper silicates pigmented acrylic coatings, have demonstrated a broad PL band for calcium copper silicate situated in the interval from $\sim 800$ to $\sim 1000$~nm.
Its maximum gradually moved from $\sim 870$ to $\sim 905$~nm as the pigment concentration increased in the acrylic film.
The band was obviously composed of a number of components that peaked in the range from $\sim 860$ to $\sim 930$~nm.

The issue of the characteristic peak position of calcium copper silicate and the changes in its position requires further investigations, which are, however, far beyond the scope of this article.

Besides the line of Egyptian blue presented in Fig.~\ref{fig:Falk_PL}\,a, the spectra plotted in Fig.~\ref{fig:Falk_PL}\,b show PL of some other pigments whose PL bands are peaked around 500~nm.

We ascribe this PL to mutually superimposed deep-level emission (DLE) bands in ZnO \cite{PL_TR_Imaging_Paintings,PL_ZnO_pigment_Zn-complexes,PL_ZnO_films_different_deposition}.
The intense cyan-green emission band of ZnO centered at $\sim 500$~nm ($\sim 2.5$~eV) is often attributed to the electron transitions from oxygen vacancies (V$_{\rm O}$) to the valence band  \cite{PL_ZnO_films_different_deposition,ZnO_films/Si(100)_RF_magnertron,PL_ZnO_V_O,ZnO_defect_emission,ZnO_hydrothermal_defects,ZnO_Review_Avrutin}.
Alternatively, it is assumed to be related to the emission of Cu$^{2+}$ ion in a zinc lattice site \cite{ZnO_Review_Avrutin}.%
\footnote{%
	Some other models of this emission are also reviewed in Ref.~\cite{ZnO_Review_Avrutin}.
}

The weak DLE bands seen in Fig.~\ref{fig:Falk_PL}\,b are also described in the literature.
The violet-blue emission band at $\sim 430$~nm ($\sim 2.85$~eV) may be associated, e.g., with the electron transition from the zinc interstitial (Zn$_i$) level to that of zinc vacancy (V$_{\rm Zn}$) \cite{PL_ZnO_films_different_deposition,ZnO_hydrothermal_defects}.
%
The DLE related blue band at $\sim 460$~nm ($\sim 2.7$~eV) has not assigned thus far. 
However, since a considerable red shift of the deep-level PL maximum from $\sim 2.7$ to $\sim 2.5$~eV was observed with the growing sample temperature up to 150~K, the authors of Ref.~\cite{PL_ZnO_films_different_deposition} supposed that the deep-level related band was composed of two peaks, blue and cyan-green, with the former one dominating at the temperature $T<150$~K and the latter one at $T>150$~K.
The greenish-yellow PL band peaking in the range from $\sim560$ to $\sim570$~nm ($\sim 2.2$~eV) is likely related to the electron transitions from the conduction band to oxygen interstitials (O$_i$),
whereas the orange band centered at 610 to 620~nm ($\sim 2$~eV) may be assigned to the Zn$_i\rightarrow$~O$_i$ electron transitions \cite{PL_ZnO_films_different_deposition}.
The nature of red PL band at $\sim680$~nm ($\sim 1.8$~eV) is unknown at present \cite{ZnO_hydrothermal_defects,ZnO_Review_Avrutin} although it is also assumed to be connected with defect-related DLE \cite{ZnO_CL}.


Note also that the emission band of hydrocerussite (2PbCO$_3${\textperiodcentered}Pb(OH)$_2$) also lies in the range 480 to 580~nm \cite{PL_TR_Imaging_Paintings}.
This band is probably superimposed on the ZnO DLE one.

Thus we can conclude that zinc and probably lead whites also contribute to the PL spectra of the samples emitting in the visible range at the wavelengths around 500~nm under B, BV, V and UV PE.

\subsubsection{A remark on the provenance of Egyptian blue in the painting by Falk}
\label{disc:EB_provenance}


We are sure that Egyptian blue detected in this painting can certainly be only of European provenance. 
The case is that, from the one hand, the manufacture of art materials was limited to only a small amount of paints of a rather poor color assortment at that period in Russia, which could not include such a rare pigment as Egyptian blue.
On the other hand, it is well known that Russian artists preferred paints produced in Europe due to their higher quality. 
Despite there is no reliable information about Egyptian blue manufacturing in Europe in the beginning of 20th century, the oil paint based on this pigment was mentioned as `Bleu de Pomp{\'{e}}i' in the Lefranc \& Cie catalogue in 1930 \cite{Lefranc-1930,Grenberg-Pisareva}.%
\footnote{%
	Additionally, the same dry pigment, Pompeian Blue, according to Ref.~\cite{Colour_Atlas}, was presented in Lefranc~\& Cie, France, in 1929.%
 } 
It can be assumed that the paint was also being produced earlier, yet, it does not seem to be popular among artists. 
It may be the cause of the lack of Egyptian blue in artworks of the 20th century.

\section{Conclusion}
\label{concl}

Summarizing the article let us present its main statements.

First, we have thoroughly analyzed the commercial Egyptian blue pigment produced by Kremer and found it suitable for utilization as a standard for PL spectral analysis.
The characteristic emission band of CaCuSi$_4$O$_{10}$, which is used for the analysis, has been found to peak at the wavelength of about 910~nm.
This band is composed of two components peaking at about 900 and 923~nm.
The PL band is efficiently excited by green or blue incoherent light.

Second, we have detected Egyptian blue pigment in the paint layer of the ``Birch. Spring'' painting by~Robert~Falk (1907, oil on canvas, private collection).
In addition, we have found the same pigment in the paints of the sketch drawn on the reverse side of the canvas. 
This is likely the first time that Egyptian blue has been discovered in 20th century paintings as well as in pieces of art of that period of time at all.

Finally, we should emphasize in conclusion that this study cogently demonstrates the micro PL spectral analysis to be an efficient tool for detecting luminescent pigments contained in paint layers of works of art.
Excitation of pigment luminescence in paints of art pieces by incoherent light is effective. 
At the same time, application of light sources enabling the PE band selection in wide ranges, which are often available in PL microscopes, considerably enhances the pigment analysis possibilities that is especially important for examining compounds, which are characterized by selective PE of luminescence and their mixtures.

\section*{Acknowledgments}

The research was accomplished under the collaboration agreement between GPI RAS and {GOSNIIR}.

\section*{Electronic supplementary material}
Figure~S1:~Photographs of pages from the Lefranc~\& Cie catalogue of 1930: (a) the cover, (b) the title page and (c) the page 10 that contains the record about Egyptian blue (``\textit{Silicate de cuivre et de chaux.} \textbf{Bleu de Pomp{\'{e}}i}'').

\paragraph{\rm\textbf{Conflict of interest}} None declared.




\begin{figure}[t]
	\begin{minipage}[l]{1\textwidth}\vspace{-4cm}\hspace{-0cm}
		\includegraphics[width=0.95\textwidth]{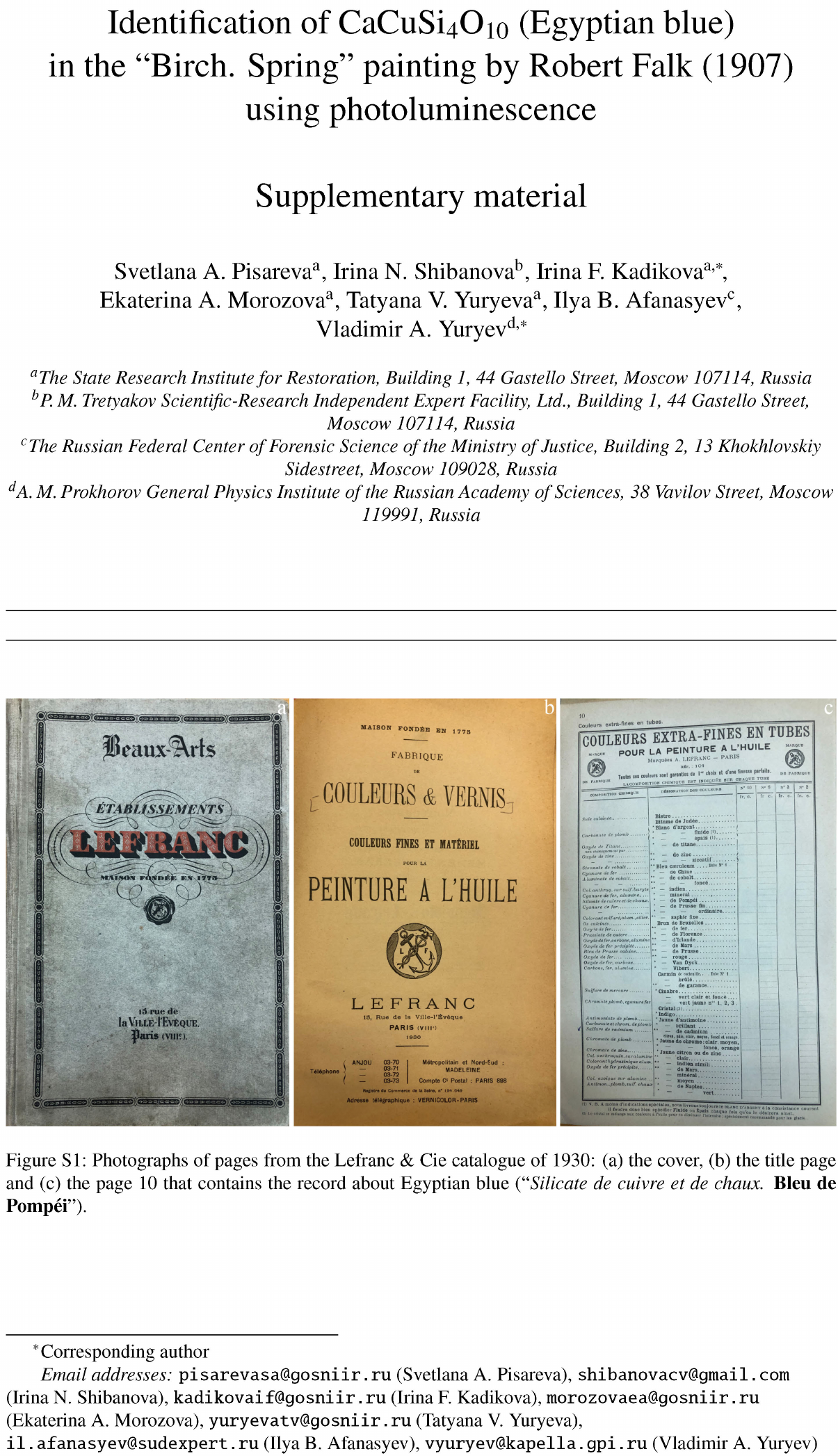}
	\end{minipage}
	\label{ESM_1-fig}
\end{figure}


\begin{thebibliography}{10}
	\expandafter\ifx\csname url\endcsname\relax
	\def\url#1{\texttt{#1}}\fi
	\expandafter\ifx\csname urlprefix\endcsname\relax\def\urlprefix{URL }\fi
	\expandafter\ifx\csname href\endcsname\relax
	\def\href#1#2{#2} \def\path#1{#1}\fi
	
	\bibitem{Pigment_Compendium}
	N.~Easthaugh, V.~Walsh, T.~Chaplin, R.~Siddall,
	\href{https://doi.org/10.4324/9780080473765}{Pigment Compendium: A Dictionary
		of Historical Pigments}, Routledge, London, UK, 2004.
	\newblock \href {https://doi.org/10.4324/9780080473765}
	{\path{doi:10.4324/9780080473765}}.
	\newline\urlprefix\url{https://doi.org/10.4324/9780080473765}
	
	\bibitem{Egyptian_Blue_book}
	J.~Riederer,
	\href{https://www.nga.gov/content/dam/ngaweb/research/publications/pdfs/artists-pigments-vol3.pdf}{Egyptian
		blue}, in: E.~W. FitzHugh (Ed.), Artists{\textquotesingle} Pigments: A
	Handbook of their History and Characteristics, Vol.~3, National Gallery of
	Art, Washington, USA, 2012, pp. 21--46.
	\newline\urlprefix\url{https://www.nga.gov/content/dam/ngaweb/research/publications/pdfs/artists-pigments-vol3.pdf}
	
	\bibitem{Ancient_Blue_Purple_Pigments}
	H.~Berke, \href{http://doi.org/10.1039/B606268G}{The invention of blue and
		purple pigments in ancient times}, Chem. Soc. Rev. 36 (2007) 15--30.
	\newblock \href {https://doi.org/10.1039/B606268G}
	{\path{doi:10.1039/B606268G}}.
	\newline\urlprefix\url{http://doi.org/10.1039/B606268G}
	
	\bibitem{Micro-EB_Arch_England}
	M.~G. Canti, J.~L. Heathcote,
	\href{http://doi.org/10.1006/jasc.2001.0717}{Microscopic {Egyptian} blue
		(synthetic cuprorivaite) from sediments at two archaeological sites in {West
			Central England}}, J. Archaeol. Sci. 29~(8) (2002) 831--836.
	\newblock \href {https://doi.org/10.1006/jasc.2001.0717}
	{\path{doi:10.1006/jasc.2001.0717}}.
	\newline\urlprefix\url{http://doi.org/10.1006/jasc.2001.0717}
	
	\bibitem{EB_Europe_Egypt-1910}
	A.~P. Laurie,
	\href{https://archive.org/details/cu31924016809927/page/n12/mode/2up}{{The
			Materials of the Painter's Craft in Europe and Egypt: from Earliest Times to
			the End of the {XVIIth} Century, with Some Account of Their Preparation and
			Use}}, T. N. Foulis, {London \& Edinburgh, UK}, 1910.
	\newline\urlprefix\url{https://archive.org/details/cu31924016809927/page/n12/mode/2up}
	
	\bibitem{Egyptian_Blue_Fayum_Portraits-2015}
	M.~Ganio, J.~Salvant, J.~Williams, L.~Lee, O.~Cossairt, M.~Walton,
	\href{https://doi.org/10.1007/s00339-015-9424-5}{Investigating the use of
		{Egyptian} blue in {Roman Egyptian} portraits and panels from {Tebtunis,
			Egypt}}, Appl. Phys. A 121~(3) (2015) 813--821.
	\newblock \href {https://doi.org/10.1007/s00339-015-9424-5}
	{\path{doi:10.1007/s00339-015-9424-5}}.
	\newline\urlprefix\url{https://doi.org/10.1007/s00339-015-9424-5}
	
	\bibitem{EB_production_technology_Mesopotamia}
	G.~D. Hatton, A.~J. Shortland, M.~S. Tite,
	\href{https://doi.org/10.1016/j.jas.2007.11.008}{The production technology of
		{Egyptian} blue and green frits from second millennium {BC Egypt and
			Mesopotamia}}, J. Archaeol. Sci. 35~(6) (2008) 1591--1604.
	\newblock \href {https://doi.org/10.1016/j.jas.2007.11.008}
	{\path{doi:10.1016/j.jas.2007.11.008}}.
	\newline\urlprefix\url{https://doi.org/10.1016/j.jas.2007.11.008}
	
	\bibitem{Colour_Atlas}
	\href{https://luciephotobookprize.com/submit/uploads/22290/51-1885-18/pdf/2bbb2c8a13987b0491d70b96f772d0cf.pdf}{An
		Atlas of Rare \& Familiar Colour. The Harvard Art Museums{\textquotesingle}
		Forbes Pigment Collection}, Atelier {\'{E}}ditions, Los Angeles, California,
	USA, 2017.
	\newline\urlprefix\url{https://luciephotobookprize.com/submit/uploads/22290/51-1885-18/pdf/2bbb2c8a13987b0491d70b96f772d0cf.pdf}
	
	\bibitem{arrigoni2015colour}
	D.~D.~T. Arrigoni, \href{https://books.google.ru/books?id=9zrLCgAAQBAJ}{Colour
		in Painting: Materials and Spirituality}, Lulu Enterprises Inc., Egham, UK,
	2015.
	\newline\urlprefix\url{https://books.google.ru/books?id=9zrLCgAAQBAJ}
	
	\bibitem{diodato2017encaustic}
	S.~P. Diodato, \href{https://books.google.ru/books?id=rIsmDwAAQBAJ}{Encaustic
		Wall Painting: How to Execute an Encaustic Painting on a Terracotta Support
		Following Recipes in Classical Sources}, Ancient and medieval art techniques.
	Handbooks, Nardini Editore, Florence, Italy, 2017.
	\newline\urlprefix\url{https://books.google.ru/books?id=rIsmDwAAQBAJ}
	
	\bibitem{Egyptian_Blue_Persia}
	E.~Aloiz, J.~G. Douglas, A.~Nagel,
	\href{https://doi.org/10.1186/s40494-016-0072-7}{Painted plaster and glazed
		brick fragments from {Achaemenid Pasargadae and Persepolis, Iran}}, Heritage
	Science 4~(1) (2016) 3.
	\newblock \href {https://doi.org/10.1186/s40494-016-0072-7}
	{\path{doi:10.1186/s40494-016-0072-7}}.
	\newline\urlprefix\url{https://doi.org/10.1186/s40494-016-0072-7}
	
	\bibitem{EB_Turkey_Ayanis}
	G.~M. Ingo, A.~{\c{C}}ilingiro{\u{g}}lu, G.~D. Carlo, A.~Batmaz, T.~D. Caro,
	C.~Riccucci, E.~I. Parisi, F.~Faraldi,
	\href{https://doi.org/10.1016/j.jas.2013.06.016}{Egyptian {Blue} cakes from
		the {Ayanis} fortress {(Eastern Anatolia, Turkey)}: micro-chemical and
		-structural investigations for the identification of manufacturing process
		and provenance}, J. Archaeol. Sci. 40~(12) (2013) 4283--4290.
	\newblock \href {https://doi.org/10.1016/j.jas.2013.06.016}
	{\path{doi:10.1016/j.jas.2013.06.016}}.
	\newline\urlprefix\url{https://doi.org/10.1016/j.jas.2013.06.016}
	
	\bibitem{EB_Turkey_Lake_Van}
	{\"{O}}.~Ormanci,
	\href{https://doi.org/10.1016/j.saa.2019.117889}{Non-destructive
		characterization of {Egyptian Blue} cakes and wall painting fragments from
		the east of {Lake Van, Turkey}}, Spectrochim. Acta A 229 (2020) 117889.
	\newblock \href {https://doi.org/10.1016/j.saa.2019.117889}
	{\path{doi:10.1016/j.saa.2019.117889}}.
	\newline\urlprefix\url{https://doi.org/10.1016/j.saa.2019.117889}
	
	\bibitem{EB_Egypt_Aegean+Near_East}
	M.~Panagiotaki, M.~S. Tite, Y.~Maniatis,
	\href{https://www.peeters-leuven.be/detail.php?search_key=9789042925502}{{Egyptian
			Blue} in {Egypt} and beyond: the {Aegean} and the {Near~East}}, in:
	P.~Kousoulis, N.~Lazaridis (Eds.), Proceedings of the Tenth International
	Congress of Egyptologists, University of the Aegean, Rhodes 22--29 May 2008,
	Vol.~{II} of Orientalia Lovaniensia Analecta 241, Peeters, Leuven -- Paris --
	Bristol, CT, 2015, pp. 1769--1790.
	\newline\urlprefix\url{https://www.peeters-leuven.be/detail.php?search_key=9789042925502}
	
	\bibitem{EB@Kos_article}
	A.~K. Marketou,
	\href{http://www.thiasos.eu/wp-content/uploads/2019/06/02-2019-Marketou_20190610.pdf}{The
		pigment production site of the ancient agora of {Kos (Greece)}: Revisiting
		the material evidence}, Thiasos 8.1 (2019) 61--80.
	\newline\urlprefix\url{http://www.thiasos.eu/wp-content/uploads/2019/06/02-2019-Marketou_20190610.pdf}
	
	\bibitem{Grenberg_&_Pisareva-Erebuni_1982}
	Y.~I. Grenberg, S.~A. Pisareva, A.~B. Levstein,
	\href{https://doi.org/10.13140/RG.2.2.25685.17129}{The study of the paint
		layer composition in the samples of mural paintings of {Erebuni}},
	Preliminary report on research project (topic 7.3), The All-Union
	Scientific-Research Institute for Restoration, Moscow, USSR, in Russian
	(1982).
	\newblock \href {https://doi.org/10.13140/RG.2.2.25685.17129}
	{\path{doi:10.13140/RG.2.2.25685.17129}}.
	\newline\urlprefix\url{https://doi.org/10.13140/RG.2.2.25685.17129}
	
	\bibitem{Pisareva-Erebuni_1987}
	S.~A. Pisareva, V.~N. Kireeva,
	\href{https://www.researchgate.net/publication/340129791_The_Study_of_the_paint_layer_composition_in_the_samples_of_mural_paintings_of_Erebuni_Issledovanie_sostava_krasocnogo_sloa_v_obrazcah_nastennyh_rospisej_iz_Erebuni}{The
		study of the paint layer composition in the samples of mural paintings of
		{Erebuni}}, in: Y.~I. Grenberg (Ed.), Investigations of monumental painting
	of Armenia and Novgorod, no.~6 in Culture and Arts in the USSR. Series:
	Restoration of Historical and Cultural Monuments. Express Information, The
	Lenin State Library of the USSR, Moscow, USSR, 1987, pp. 1--2, in Russian.
	\newline\urlprefix\url{https://www.researchgate.net/publication/340129791_The_Study_of_the_paint_layer_composition_in_the_samples_of_mural_paintings_of_Erebuni_Issledovanie_sostava_krasocnogo_sloa_v_obrazcah_nastennyh_rospisej_iz_Erebuni}
	
	\bibitem{Egyptian_Blue_Visible-Induced}
	R.~Linn, E.~H. Cline, A.~Yasur-Landau,
	\href{https://www.sidestone.com/books/tracing-technoscapes}{The advantages of
		visible induced luminescence technique for the investigation of
		{Aegean-style} wall painting. {A} case study from {Tel Kabri, Israel}}, in:
	J.~Becker, J.~Jungfleisch, C.~von R{\"{u}}den (Eds.), Tracing Technoscapes.
	The Production of Bronze Age Wall Paintings in the {Eastern Mediterranean},
	Sidestone Press, Leiden, The Netherlands, 2018, pp. 101--116.
	\newline\urlprefix\url{https://www.sidestone.com/books/tracing-technoscapes}
	
	\bibitem{Egyptian_Blue_Fayum_Portraits-2018}
	J.~Salvant, J.~Williams, M.~Ganio, F.~Casadio, C.~Daher, K.~Sutherland,
	L.~Monico, F.~Vanmeert, S.~De~Meyer, K.~Janssens, C.~Cartwright, M.~Walton,
	\href{https://doi.org/10.1111/arcm.12351}{A {Roman Egyptian} painting
		workshop: Technical investigation of the portraits from {Tebtunis, Egypt}},
	Archaeometry 60~(4) (2018) 815--833.
	\newblock \href {https://doi.org/10.1111/arcm.12351}
	{\path{doi:10.1111/arcm.12351}}.
	\newline\urlprefix\url{https://doi.org/10.1111/arcm.12351}
	
	\bibitem{Old_Nisa_Veresotskaya}
	G.~E. Veresotskaya,
	\href{http://www.gosniir.ru/library/artistic-heritage/artistic-heritage-23.aspx}{Restoration
		of a fragment of the {``Horsemen''} archaeological painting from {Old Nisa}},
	in: Artistic Heritage. Conservation, Research, Restoration, no. 23(53), The
	State Research Institute for Restoration, Moscow, Russia, 2006, pp. 166--174,
	190, 191, {Also in}: ARTconservation, https://web.archive.org/web/20121111061508/http://art-con.ru/node/1222.
	\newline\urlprefix\url{http://www.gosniir.ru/library/artistic-heritage/artistic-heritage-23.aspx}
	
	\bibitem{EB_icons_Naumova-1983}
	M.~M. Naumova, V.~Y. Birshtein,
	\href{http://www.gosniir.ru/library/artistic-heritage/artistic-heritage-08.aspx}{Comparison
		of the paint layers characteristics and painting materials of three encaustic
		icons from {Kiev Museum} of {Eastern and Western Art}}, in: Artistic
	Heritage. Conservation, Research, Restoration, no. 8(38), The All-Union
	Research Institute for Restoration (VNIIR), Moscow, USSR, 1983, pp. 150--153,
	in Russian.
	\newline\urlprefix\url{http://www.gosniir.ru/library/artistic-heritage/artistic-heritage-08.aspx}
	
	\bibitem{EB_Norway}
	A.~M. Rosenqvist,
	\href{https://www.duo.uio.no/bitstream/handle/10852/37585/1959-vol-23.pdf}{Analyser
		an skerd og skjold fra {B{\o}}-funnet ({Analysis} of sword and shield from
		the {B{\o}}-find)}, Viking 23 (1959) 29--34.
	\newline\urlprefix\url{https://www.duo.uio.no/bitstream/handle/10852/37585/1959-vol-23.pdf}
	
	\bibitem{EB_Zn-rich_Nicola2019}
	M.~Nicola, L.~M. Seymour, M.~Aceto, E.~Priola, R.~Gobetto, A.~Masic,
	\href{https://doi.org/10.1007/s12520-019-00873-w}{Late production of
		{Egyptian} blue: synthesis from brass and its characteristics}, Archaeol.
	Anthrop. Sci. 11~(10) (2019) 5377--5392.
	\newblock \href {https://doi.org/10.1007/s12520-019-00873-w}
	{\path{doi:10.1007/s12520-019-00873-w}}.
	\newline\urlprefix\url{https://doi.org/10.1007/s12520-019-00873-w}
	
	\bibitem{EB_Zn-rich_Castelseprio_NICOLA2018465}
	M.~Nicola, M.~Aceto, V.~Gheroldi, R.~Gobetto, G.~Chiari,
	\href{https://doi.org/10.1016/j.jasrep.2018.03.031}{Egyptian blue in the
		{Castelseprio} mural painting cycle. {Imaging} and evidence of a
		non-traditional manufacture}, J. Archaeol. Sci. Rep. 19 (2018) 465--475.
	\newblock \href {https://doi.org/10.1016/j.jasrep.2018.03.031}
	{\path{doi:10.1016/j.jasrep.2018.03.031}}.
	\newline\urlprefix\url{https://doi.org/10.1016/j.jasrep.2018.03.031}
	
	\bibitem{EB_Spain}
	A.~Lluveras, A.~Torrents, P.~Gir{\'{a}}lez, M.~Vendrell-Saz,
	\href{https://doi.org//10.1111/j.1475-4754.2009.00481.x}{Evidence for the use
		of {Egyptian} blue in an 11th century mural altarpiece by {SEM--EDS}, {FTIR}
		and {SR~XRD} {(church of Sant Pere, Terrassa, Spain)}}, Archaeometry 52~(2)
	(2010) 308--319.
	\newblock \href {https://doi.org/10.1111/j.1475-4754.2009.00481.x}
	{\path{doi:10.1111/j.1475-4754.2009.00481.x}}.
	\newline\urlprefix\url{https://doi.org//10.1111/j.1475-4754.2009.00481.x}
	
	\bibitem{Egyptian_Blue_Benvenuto-1524}
	J.~Bredal-J{\o}rgensen, J.~Sanyova, V.~Rask, M.~L. Sargent, R.~H. Therkildsen,
	\href{https://doi.org/10.1007/s00216-011-5140-y}{Striking presence of
		{Egyptian} blue identified in a painting by {Giovanni Battista Benvenuto}
		from 1524}, Anal. Bioanal. Chem. 401~(4) (2011) 1433--1439.
	\newblock \href {https://doi.org/10.1007/s00216-011-5140-y}
	{\path{doi:10.1007/s00216-011-5140-y}}.
	\newline\urlprefix\url{https://doi.org/10.1007/s00216-011-5140-y}
	
	\bibitem{Cuprorivaite_discovery}
	C.~Minguzzi, Cuprorivaite: {Un} nuovo minerale, Period. Mineral. 9~(3) (1938)
	333--345.
	
	\bibitem{Pabst:cuprorivaite}
	A.~Pabst, \href{https://doi.org/10.1107/S0365110X5900216X}{{Structures of some
			tetragonal sheet silicates}}, Acta Cryst. 12~(10) (1959) 733--739.
	\newblock \href {https://doi.org/10.1107/S0365110X5900216X}
	{\path{doi:10.1107/S0365110X5900216X}}.
	\newline\urlprefix\url{https://doi.org/10.1107/S0365110X5900216X}
	
	\bibitem{Reexamination_cuprorivaite}
	F.~Mazzi, A.~Pabst,
	\href{https://www.minsocam.org/ammin/AM47/AM47_409.pdf}{Reexamination of
		cuprorivaite}, Am. Min. 47 (1962) 409--411.
	\newline\urlprefix\url{http://www.minsocam.org/ammin/AM47/AM47_409.pdf}
	
	\bibitem{Cuprorivaite_Structure}
	B.~C. Chakoumakos, J.~A. Fernandez-Baca, L.~A. Boatner,
	\href{http://doi.org/10.1006/jssc.1993.1083}{Refinement of the structures of
		the layer silicates {MCuSi$_4$O$_{10}$ (M = Ca, Sr, Ba)} by {Rietveld}
		analysis of neutron powder diffraction data}, J. Solid State Chem. 103~(1)
	(1993) 105--113.
	\newblock \href {https://doi.org/10.1006/jssc.1993.1083}
	{\path{doi:10.1006/jssc.1993.1083}}.
	\newline\urlprefix\url{http://doi.org/10.1006/jssc.1993.1083}
	
	\bibitem{Bensch:cuprorivaite}
	W.~Bensch, M.~Schur, \href{https://doi.org/10.1524/zkri.1995.210.7.530}{Crystal
		structure of calcium copper phyllo-decaoxotetrasilicate,
		{CaCuSi$_4$O$_{10}$}}, Z. Kristallogr. Cryst. Mater. 210~(7) (1995) 530--530.
	\newblock \href {https://doi.org/10.1524/zkri.1995.210.7.530}
	{\path{doi:10.1524/zkri.1995.210.7.530}}.
	\newline\urlprefix\url{https://doi.org/10.1524/zkri.1995.210.7.530}
	
	\bibitem{Cuprorivaite}
	\href{http://www.handbookofmineralogy.org/pdfs/cuprorivaite.pdf}{Cuprorivaite.
		{CaCuSi$_4$O$_{10}$}}, in: J.~W. Anthony, R.~A. Bideaux, K.~W. Bladh, M.~C.
	Nichols (Eds.), Handbook of Mineralogy, Vol.~II, Mineralogical Society of
	America, Chantilly, VA, USA, 2003.
	\newline\urlprefix\url{http://www.handbookofmineralogy.org/pdfs/cuprorivaite.pdf}
	
	\bibitem{Cuprorivaite_mindata}
	\href{https://www.mindat.org/min-1189.html}{Cuprorivaite}, Hudson Institute of
	Mineralogy, Keswick, VA, USA, dba Mindat.org (2019).
	\newline\urlprefix\url{https://www.mindat.org/min-1189.html}
	
	\bibitem{Egyptian_Blue_PL}
	G.~Accorsi, G.~Verri, M.~Bolognesi, N.~Armaroli, C.~Clementi, C.~Miliani,
	A.~Romani, \href{https://doi.org/10.1039/B902563D}{The exceptional
		near-infrared luminescence properties of cuprorivaite {(Egyptian blue)}},
	Chem. Commun. 23 (2009) 3392--3394.
	\newblock \href {https://doi.org/10.1039/B902563D}
	{\path{doi:10.1039/B902563D}}.
	\newline\urlprefix\url{https://doi.org/10.1039/B902563D}
	
	\bibitem{PL_some_Blue_natural_pigments}
	D.~Aj\`{o}, G.~Chiari, F.~Zuane, M.~Favaro, M.~Bertolin,
	\href{https://www.researchgate.net/publication/317953199_Photoluminescence_of_some_Blue_natural_pigments_and_related_synthetic_materials}{Photoluminescence
		of some {Blue} natural pigments and related synthetic materials}, in:
	Non-Destructive Testing, Budapest, Hungary, 1996.
	\newline\urlprefix\url{https://www.researchgate.net/publication/317953199_Photoluminescence_of_some_Blue_natural_pigments_and_related_synthetic_materials}
	
	\bibitem{EB-PL_Application_Triolo2019}
	P.~A.~M. Triolo, M.~Spingardi, G.~A. Costa, F.~Locardi,
	\href{https://doi.org/10.1007/s12520-019-00848-x}{Practical application of
		visible-induced luminescence and use of parasitic {IR} reflectance as
		relative spatial reference in {Egyptian} artifacts}, Archaeol. Anthrop. Sci.
	11~(9) (2019) 5001--5008.
	\newblock \href {https://doi.org/10.1007/s12520-019-00848-x}
	{\path{doi:10.1007/s12520-019-00848-x}}.
	\newline\urlprefix\url{https://doi.org/10.1007/s12520-019-00848-x}
	
	\bibitem{MCuSi4O10_IR_phosphors}
	P.~Berdahl, S.~K. Boocock, G.~C.-Y. Chan, S.~S. Chen, R.~M. Levinson, M.~A.
	Zalich, \href{https://doi.org/10.1063/1.5019808}{High quantum yield of the
		{Egyptian} blue family of infrared phosphors {(MCuSi$_4$O$_{10}$, M = Ca, Sr,
			Ba)}}, J. Appl. Phys. 123~(19) (2018) 193103.
	\newblock \href {https://doi.org/10.1063/1.5019808}
	{\path{doi:10.1063/1.5019808}}.
	\newline\urlprefix\url{https://doi.org/10.1063/1.5019808}
	
	\bibitem{EB_NIR_emission_upconversion}
	W.~Chen, Y.~Shi, Z.~Chen, X.~Sang, S.~Zheng, X.~Liu, J.~Qiu,
	\href{https://doi.org/10.1021/acs.jpcc.5b04819}{Near-infrared emission and
		photon energy upconversion of two-dimensional copper silicates}, J. Phys.
	Chem. C 119~(35) (2015) 20571--20577.
	\newblock \href {https://doi.org/10.1021/acs.jpcc.5b04819}
	{\path{doi:10.1021/acs.jpcc.5b04819}}.
	\newline\urlprefix\url{https://doi.org/10.1021/acs.jpcc.5b04819}
	
	\bibitem{PL_EB_HB_HP}
	G.~Pozza, D.~Aj\`{o}, G.~Chiari, F.~D. Zuane, M.~Favaro,
	\href{https://doi.org/10.1016/S1296-2074(00)01095-5}{Photoluminescence of the
		inorganic pigments {Egyptian} blue, {Han} blue and {Han} purple}, J. Cult.
	Herit. 1~(4) (2000) 393--398.
	\newblock \href {https://doi.org/https://10.1016/S1296-2074(00)01095-5}
	{\path{doi:10.1016/S1296-2074(00)01095-5}}.
	\newline\urlprefix\url{https://doi.org/10.1016/S1296-2074(00)01095-5}
	
	\bibitem{Egyptian_Blue_New_Life}
	S.~M. Borisov, C.~W{\"{u}}rth, U.~Resch-Genger, I.~Klimant,
	\href{https://doi.org/10.1021/ac402275g}{New life of ancient pigments:
		Application in high-performance optical sensing materials}, Anal. Chem.
	85~(19) (2013) 9371--9377.
	\newblock \href {https://doi.org/10.1021/ac402275g}
	{\path{doi:10.1021/ac402275g}}.
	\newline\urlprefix\url{https://doi.org/10.1021/ac402275g}
	
	\bibitem{Warner_EB_Synthesis-2011}
	T.~E. Warner, Artificial cuprorivaite {CaCuSi$_4$O$_{10}$ (Egyptian blue)} by a
	salt-flux method, in: Synthesis, Properties and Mineralogy of Important
	Inorganic Materials, Wiley, UK, 2011, Ch.~3, pp. 26--49.
	
	\bibitem{Egyptian_Blue_Nano}
	D.~Johnson-McDaniel, C.~A. Barrett, A.~Sharafi, T.~T. Salguero,
	\href{https://doi.org/10.1021/ja310587c}{Nanoscience of an ancient pigment},
	J. Am. Chem. Soc. 135 (2013) 1677--1679.
	\newblock \href {https://doi.org/10.1021/ja310587c}
	{\path{doi:10.1021/ja310587c}}.
	\newline\urlprefix\url{https://doi.org/10.1021/ja310587c}
	
	\bibitem{EB_hydrothermal}
	D.~Johnson-McDaniel, S.~Comer, J.~W. Kolis, T.~T. Salguero,
	\href{https://doi.org/10.1002/chem.201503364}{Hydrothermal formation of
		calcium copper tetrasilicate}, Chem. Eur. J. 21~(49) (2015)
	17560--17564.
	\newblock \href {https://doi.org/10.1002/chem.201503364}
	{\path{doi:10.1002/chem.201503364}}.
	\newline\urlprefix\url{https://doi.org/10.1002/chem.201503364}
	
	\bibitem{EB_Exfoliation}
	D.~Johnson-McDaniel, T.~T. Salguero,
	\href{http://www.jove.com/video/51686}{Exfoliation of {Egyptian} blue and
		{Han} blue, two alkali earth copper silicate-based pigments}, J. Vis.
	Exp.~(86) (2014) e51686.
	\newblock \href {https://doi.org/10.3791/51686} {\path{doi:10.3791/51686}}.
	\newline\urlprefix\url{http://www.jove.com/video/51686}
	
	\bibitem{EB_ML_structure&properties}
	Y.~Chen, M.~Kan, Q.~Sun, P.~Jena,
	\href{https://doi.org/10.1021/acs.jpclett.5b02770}{Structure and properties
		of egyptian blue monolayer family: {XCuSi$_4$O$_{10}$} {(X = Ca, Sr, and
			Ba)}}, J. Phys. Chem. Lett. 7~(3) (2016) 399--405.
	\newblock \href {https://doi.org/10.1021/acs.jpclett.5b02770}
	{\path{doi:10.1021/acs.jpclett.5b02770}}.
	\newline\urlprefix\url{https://doi.org/10.1021/acs.jpclett.5b02770}
	
	\bibitem{EB_fingerprint_powder}
	B.~Errington, G.~Lawson, S.~W. Lewis, G.~D. Smith,
	\href{https://doi.org/10.1016/j.dyepig.2016.05.008}{Micronised egyptian blue
		pigment: A novel near-infrared luminescent fingerprint dusting powder}, Dyes
	Pigm. 132 (2016) 310--315.
	\newblock \href {https://doi.org/10.1016/j.dyepig.2016.05.008}
	{\path{doi:10.1016/j.dyepig.2016.05.008}}.
	\newline\urlprefix\url{https://doi.org/10.1016/j.dyepig.2016.05.008}
	
	\bibitem{R.Falk_Pisareva}
	I.~N. Shibanova, S.~A. Pisareva, Y.~V. Chaykina,
	\href{https://www.crys.ras.ru/document/RCEM-2018/RCEM-2018_abstracts_volume_2_(conference).pdf}{Electron
		microscopy as one of the methods in the examination of paintings}, in: The
	{XXVII} Russian Conference ``Modern Methods of Electron and Probe Microscopy
	in Studies of Organic, Inorganic Nanostructures and Nano-Biomaterials''
	(RCEM--2018), August 28--30, 2018, Chernogolovka, Moscow Region, Russia, Book
	of Abstracts, Vol.~2, FNIC ``Crystallography and Photonics'', Moscow, Russia,
	2018, pp. 114--115, in Russian.
	\newline\urlprefix\url{https://www.crys.ras.ru/document/RCEM-2018/RCEM-2018_abstracts_volume_2_(conference).pdf}
	
	\bibitem{Falk_Besancon}
	A.~Besan{\c{c}}on, \href{https://doi.org/10.3406/cmr.1962.1527}{{R. R. Falk}
		(1886\,--\,1958)}, Cahiers du Monde Russe et Sovi{\'{e}}tique 3~(4) (1962)
	564--581.
	\newblock \href {https://doi.org/10.3406/cmr.1962.1527}
	{\path{doi:10.3406/cmr.1962.1527}}.
	\newline\urlprefix\url{https://doi.org/10.3406/cmr.1962.1527}
	
	\bibitem{Falk_Sarabjanow}
	D.~Sarabjanow, {Robert Falk. Mit einer Dokumentation, Briefen,
		Gespr{\"{a}}chen, Lektionen des K{\"{u}}nstlers und einer biographischen
		{\"{U}}bersicht, herausgegeben von A.~W.~Stschekin-Krotowa}, Verlag der
	Kunst, Dresden, DDR, 1974, in German.
	
	\bibitem{Falk_Stschekin-Krotowa}
	R.~R. Falk, A.~V. {Shchekin-Krotova~(ed.)}, Conversations about art. Letters.
	Memories about the artist, Soviet Artist, Moscow, USSR, 1981, in Russian.
	
	\bibitem{My_Falk_Stschekin-Krotowa}
	A.~V. Shchekin-Krotova, My Falk, HGS, Moscow, Russia, 2005, in Russian.
	
	\bibitem{Falk_catalog}
	D.~V. Sarabyanov, Y.~I. Didenko,
	\href{https://books.google.com/books?id=RGVHAQAAIAAJ}{Painting by Robert
		Falk: A complete catalog of works}, Computer Press, Moscow, Russia, 2006, in
	Russian.
	\newline\urlprefix\url{https://books.google.com/books?id=RGVHAQAAIAAJ}
	
	\bibitem{X-ray_Diffraction_Book}
	B.~E. Warren, X-ray Diffraction, Addison--Wesley, Reading, MA/Dover, Mineola,
	NY, USA, 1969.
	
	\bibitem{commercial_pigments_reliable}
	L.~Rampazzi, C.~Corti,
	\href{http://www.ijcs.uaic.ro/public/IJCS-19-19_Rampazzi.pdf}{Are commercial
		pigments reliable references for the analysis of paintings?}, Int. J.
	Conserv. Sci. 10~(2) (2019) 207--220.
	\newline\urlprefix\url{http://www.ijcs.uaic.ro/public/IJCS-19-19_Rampazzi.pdf}
	
	\bibitem{Pigment_Compendium_Optical}
	N.~Easthaugh, V.~Walsh, T.~Chaplin, R.~Siddall,
	\href{https://doi.org/10.4324/9780080454573}{Pigment Compendium: Optical
		Microscopy of Historical Pigments}, Routledge, London, UK, 2005.
	\newblock \href {https://doi.org/10.4324/9780080454573}
	{\path{doi:10.4324/9780080454573}}.
	\newline\urlprefix\url{https://doi.org/10.4324/9780080454573}
	
	\bibitem{Artists_Pigments_book_PM}
	R.~L. Feller, M.~Bayard,
	\href{https://www.nga.gov/content/dam/ngaweb/research/publications/pdfs/artists-pigments-vol1.pdf}{Terminology
		and procedures used in the systematic examination of pigment particles with
		the polarizing microscope}, in: R.~L. Feller (Ed.), Artists{\textquotesingle}
	Pigments: A Handbook of their History and Characteristics, Vol.~1, National
	Gallery of Art, Washington, USA, 1986, pp. 285--298.
	\newline\urlprefix\url{https://www.nga.gov/content/dam/ngaweb/research/publications/pdfs/artists-pigments-vol1.pdf}
	
	\bibitem{PL_TR_Imaging_Paintings}
	M.~Ghirardello, G.~Valentini, L.~Toniolo, R.~Alberti, M.~Gironda, D.~Comelli,
	\href{https://doi.org/10.1016/j.microc.2020.104618}{Photoluminescence imaging
		of modern paintings: there is plenty of information at the microsecond
		timescale}, Microchem. J. 154 (2020) 104618.
	\newblock \href {https://doi.org/10.1016/j.microc.2020.104618}
	{\path{doi:10.1016/j.microc.2020.104618}}.
	\newline\urlprefix\url{https://doi.org/10.1016/j.microc.2020.104618}
	
	\bibitem{PL_ZnO_pigment_Zn-complexes}
	A.~Artesani, F.~Gherardi, A.~Nevin, G.~Valentini, D.~A. Comelli,
	\href{https://doi.org/10.3390/ma10040340}{Photoluminescence study of the
		changes induced in the zinc white pigment by formation of zinc complexes},
	Materials 10~(4) (2017) 340.
	\newblock \href {https://doi.org/10.3390/ma10040340}
	{\path{doi:10.3390/ma10040340}}.
	\newline\urlprefix\url{https://doi.org/10.3390/ma10040340}
	
	\bibitem{PL_ZnO_films_different_deposition}
	C.~H. Ahn, Y.~Y. Kim, D.~C. Kim, S.~K. Mohanta, H.~K. Cho,
	\href{https://doi.org/10.1063/1.3054175}{A comparative analysis of deep level
		emission in {ZnO} layers deposited by various methods}, J. Appl. Phys.
	105~(1) (2009) 013502.
	\newblock \href {https://doi.org/10.1063/1.3054175}
	{\path{doi:10.1063/1.3054175}}.
	\newline\urlprefix\url{https://doi.org/10.1063/1.3054175}
	
	\bibitem{ZnO_films/Si(100)_RF_magnertron}
	S.-H. Jeong, B.-S. Kim, B.-T. Lee,
	\href{https://doi.org/10.1063/1.1568543}{Photoluminescence dependence of
		{ZnO} films grown on {Si(100)} by radio-frequency magnetron sputtering on the
		growth ambient}, Appl. Phys. Lett. 82~(16) (2003) 2625--2627.
	\newblock \href {https://doi.org/10.1063/1.1568543}
	{\path{doi:10.1063/1.1568543}}.
	\newline\urlprefix\url{https://doi.org/10.1063/1.1568543}
	
	\bibitem{PL_ZnO_V_O}
	F.~K. Shan, G.~X. Liu, W.~J. Lee, B.~C. Shin,
	\href{https://doi.org/10.1063/1.2437122}{The role of oxygen vacancies in
		epitaxial-deposited {ZnO} thin films}, J. Appl. Phys. 101~(5) (2007) 053106.
	\newblock \href {https://doi.org/10.1063/1.2437122}
	{\path{doi:10.1063/1.2437122}}.
	\newline\urlprefix\url{https://doi.org/10.1063/1.2437122}
	
	\bibitem{ZnO_defect_emission}
	A.~B. Djuri{\v{s}}i{\'{c}}, Y.~H. Leung, K.~H. Tam, Y.~F. Hsu, L.~Ding, W.~K.
	Ge, Y.~C. Zhong, K.~S. Wong, W.~K. Chan, H.~L. Tam, K.~W. Cheah, W.~M. Kwok,
	D.~L. Phillips, \href{{https://doi.org/10.1088/0957-4484/18/9/095702}}{Defect
		emissions in {ZnO} nanostructures}, Nanotechnology 18~(9) (2007) 095702.
	\newblock \href {https://doi.org/10.1088/0957-4484/18/9/095702}
	{\path{doi:10.1088/0957-4484/18/9/095702}}.
	\newline\urlprefix\url{{https://doi.org/10.1088/0957-4484/18/9/095702}}
	
	\bibitem{ZnO_hydrothermal_defects}
	K.~H. Tam, C.~K. Cheung, Y.~H. Leung, A.~B. Djuri{\v{s}}i{\'{c}}, C.~C. Ling,
	C.~D. Beling, S.~Fung, W.~M. Kwok, W.~K. Chan, D.~L. Phillips, L.~Ding, W.~K.
	Ge, \href{https://doi.org/10.1021/jp063239w}{Defects in {ZnO} nanorods
		prepared by a hydrothermal method}, J. Phys. Chem. B 110~(42) (2006)
	20865--20871.
	\newblock \href {https://doi.org/10.1021/jp063239w}
	{\path{doi:10.1021/jp063239w}}.
	\newline\urlprefix\url{https://doi.org/10.1021/jp063239w}
	
	\bibitem{ZnO_Review_Avrutin}
	{\"{U}}.~{\"{O}}zg{\"{u}}r, Y.~I. Alivov, C.~Liu, A.~Teke, M.~A. Reshchikov,
	S.~Do{\u{g}}an, V.~Avrutin, S.-J. Cho, H.~Morko{\c{c}},
	\href{https://doi.org/10.1063/1.1992666}{A comprehensive review of {ZnO}
		materials and devices}, J. Appl. Phys. 98~(4) (2005) 041301.
	\newblock \href {https://doi.org/10.1063/1.1992666}
	{\path{doi:10.1063/1.1992666}}.
	\newline\urlprefix\url{https://doi.org/10.1063/1.1992666}
	
	\bibitem{ZnO_CL}
	A.~{El Hichou}, M.~Addou, J.~Eboth{\'{e}}, M.~Troyon,
	\href{https://doi.org/10.1016/j.jlumin.2004.09.123}{Influence of deposition
		temperature {($T_{\rm s}$)}, air flow rate ($f$) and precursors on
		cathodoluminescence properties of {ZnO} thin films prepared by spray
		pyrolysis}, J. Lumin. 113~(3) (2005) 183--190.
	\newblock \href {https://doi.org/10.1016/j.jlumin.2004.09.123}
	{\path{doi:10.1016/j.jlumin.2004.09.123}}.
	\newline\urlprefix\url{https://doi.org/10.1016/j.jlumin.2004.09.123}
	
	\bibitem{Lefranc-1930}
	Fabrique de Couleurs {\&} Vernis. Couleurs Fines et Mat{\'{e}}riel pour la
	Peinture a l'Huile, Lefranc, Paris, France, 1930, Ch. Couleurs extra-fines en
	tubes, p.~10.
	
	\bibitem{Grenberg-Pisareva}
	Y.~I. Grenberg, S.~A. Pisareva, Oil Paints of the {XX} Century and Examination
	of Works of Art. Composition, Disclosure, Commercial Production and
	Investigation of Paints, Zerkalo Mira, Moscow, Russia, 2010, in Russian.
	
\end{thebibliography}

\end{document}